\pdfoutput = 1
\documentclass[final,oneeqnum,onefignum]{siamltex}
\usepackage{graphicx}
\usepackage{amssymb}
\usepackage{amsmath}
\usepackage{tikz}
\usepackage{pgf}
\usepackage{pgfplots} \pgfplotsset{compat=newest}
\usepgfplotslibrary{groupplots}
\usepgfplotslibrary{external}
\usepackage[numbers,sort&compress]{natbib}

\renewcommand{\exp}[1]{e^{#1}}
\renewcommand{\vec}[1]{\mathbf{#1}}

\newcommand{\order}{\mathcal{O}}
\newcommand{\Lop}{\mathcal{L}}
\newcommand{\Qop}{\mathcal{Q}}

\title{Biological aggregation driven by social and environmental factors: A nonlocal model and its degenerate Cahn-Hilliard approximation}
\author{Andrew J. Bernoff \footnotemark[2]\ \footnotemark[4] \and Chad M. Topaz\footnotemark[3]\ \footnotemark[5]\ \footnotemark[6]}

\begin{document}

\maketitle
\renewcommand{\thefootnote}{\fnsymbol{footnote}}
\footnotetext[2]{Department of Mathematics, Harvey Mudd College, Claremont, California 91711 USA}
\footnotetext[3]{Department of Mathematics, Statistics, and Computer Science, Macalester College, St. Paul, Minnesota 55105 USA}
\footnotetext[4]{ajb@hmc.edu}
\footnotetext[5]{ctopaz@macalester.edu}
\footnotetext[6]{Supported by National Science Foundation grants DMS-1009633 and DMS-1412674 to CMT and Simons Foundation grant no. 317319 to AJB.}
\renewcommand{\thefootnote}{\arabic{footnote}}

\begin{abstract}
Biological aggregations such as insect swarms and bird flocks may arise from a combination of social interactions and environmental cues. We focus on nonlocal continuum equations, which are often used to model aggregations, and yet which pose significant analytical and computational challenges. Beginning with a particular nonlocal aggregation model [Topaz \emph{et al.}, Bull. Math. Bio., 2006], we derive the minimal well-posed long-wave approximation, which is a degenerate Cahn-Hilliard equation. Energy minimizers of this reduced, local model retain many salient features of those of the nonlocal model, especially for large populations and away from an aggregation's boundaries. Using the Cahn-Hilliard model as a testbed, we investigate the degree to which an external potential modeling food sources can be used to suppress peak population density, which is essential for controlling locust outbreaks. A random distribution of food sources tends to increase peak density above its intrinsic value, while a periodic pattern of food sources can decrease it.
\end{abstract}


\pagestyle{myheadings}
\thispagestyle{plain}
\markboth{A.J. BERNOFF AND C.M. TOPAZ}{NONLOCAL AND CAHN-HILLIARD AGGREGATION MODELS}

\section{Introduction}
\label{sec:intro}

This paper has three primary aims. First, beginning with a partial integrodifferential equation model of biological aggregation, we show that a long-wave approximation yields a degenerate Cahn-Hilliard equation akin to models describing phase separation in materials science and evolution of thin fluid films. Second, we study solutions of the degenerate Cahn-Hillard model and demonstrate that they compare well with those of the original model. Using the Cahn-Hilliard model is advantageous because it eliminates many of the analytical and computational challenges of nonlocal equations, particularly in higher dimensions. Our methodology is to study the energy minimizers of this local equation, rather than studying the time evolution of the original nonlocal equation. Finally, we use the Cahn-Hilliard model as a testbed to assess whether imposing an external potential modeling the environment, \emph{e.g.}, food sources, can be used to control peak density in aggregations, which is of interest for locust swarms.

Biological aggregations such as bird flocks, fish schools, and insect swarms are driven by social interactions between group members. Typical social forces include attraction, repulsion, and alignment \cite{CouKraJam2002,EftVriLew2007} which are preprogrammed by evolution and which are facilitated by sight, sound, smell, touch, or some combination of these senses. The accepted biological paradigm is that repulsion operates over shorter distances than attraction, but that it is much stronger over those distances. In the context of many mathematical models, these are necessary conditions to achieve cohesive groups with finite densities \cite{MogEdeBen2003,LevTopBer2009}. 

Because sensing operates over finite rather than infinitesimal distances, it is most naturally described via nonlocal operators. The most common mathematical descriptions assume that social interactions take place in an additive pairwise manner, and that the strength of interaction of organisms depends on the distance between them. We will use a kinematic model, where an individual's velocity is determined by social interactions dependent on the positions of other organisms; we will eventually augment this model to allow taxis due to environmental effects. 

For a population that is well-described by a continuum density $u(\vec{x},t)$, the social velocity $\vec{v}$, can be modeled as
\begin{equation}
\vec{v} (\vec{x},t) =\int \vec{f}(\vec{x}-\vec{y})u(\vec{y},t)\,d\vec{y} \equiv \vec{f} * u.
\end{equation}
Because the interactions are pairwise, additive and solely a function of position, the velocity can be written as the convolution of a social kernel, $\vec{f}$, that encodes the influence of organisms at location $\vec{y}$ on those at location $\vec{x}$. Such convolution-type continuum aggregation models date back to the seminal work of \cite{Kaw1978}, which examines
\begin{equation}
\label{eq:kawasaki}
\dot{u} + \nabla \cdot (u \vec{v}) = D \nabla^2 u , \quad \vec{v} =\vec{f} * u,
\end{equation}
in one spatial dimension. The left hand side of this equation is the material derivative of the population density $u$, that is, the  rate of change of the  population density moving at the velocity arising from social interactions. On the right hand side, $D$ is a diffusion constant. The diffusion term was intended to model social repulsion operating over short distances, driving flux down population gradients. This model conserves mass, preserves positivity of the density, and can display an instability from a homogeneous state to patterned states. However, a key feature of biological swarms is that they are compactly supported with sharp edges, and (\ref{eq:kawasaki}) cannot reproduce this property. This is due to the linear diffusion; even compactly supported initial state are instantaneously smoothed and densities become positive throughout the domain.

More recently, a population model that has received much attention is the aggregation equation,
\begin{equation}
\label{eq:nonlocal-intro}
\dot{u} + \nabla \cdot (u \vec{v}) =0 , \quad \vec{v} =-\nabla Q * u,
\end{equation}
where the velocity, $\vec{v}$, can incorporate both attractive and repulsive social forces. The social kernel is written as minus the gradient of a social potential, $-\nabla Q$.  One typically assumes that the social potential is directionally isotropic and symmetric (the social force between two organisms is equal in magnitude and opposite in orientation) which allows us to write the social potential as $Q=Q(r)$, solely a function of the distance between organisms, $r$. Typically $Q(r)$ is decreasing for small $r$ to model repulsion, increasing over larger $r$ to model attraction, and asymptotically flat for large $r$ as organisms cannot sense each other at large distances.

The aggregation equation has recently been the subject of a rich and rapidly growing literature, including \cite{BodVel2005,BodVel2006,BerLau2007,BerCarLau2009,BerLau2009,BerBra2010,HuaBer2010,BerLauRos2011,FetHuaKol2011,HuaBer2012,BerLauLeg2012,BerGarLau2012,HuaWitBer2012,SunUmiBer2012,BreBer2012,FetHua2013} and has also been adapted to describe specific biological organisms, such as locusts \cite{TopDOrEde2012}. Crucially, (\ref{eq:nonlocal-intro}) minimizes an energy,
\begin{equation}
\label{eq:energy-agg-intro}
E(u) = \frac 12 \int \int u(\vec{x}) Q(|\vec{x}-\vec{y}|) u(\vec{y}) \, d\vec{y} \,d\vec{x},
\end{equation}
which can be used to analyze the existence and stability of equilibria \cite{BerTop2011,BerTop2013,FetHua2013,CarChiHua2014,SimSleTop2015}. Despite its popularity in the literature, one shortfall of this model is that it does not produce so-called well-spaced groups. In well-spaced biological swarms, individuals tend to pack only up to a maximum density, and no further. In contrast, for (\ref{eq:nonlocal-intro}), if there is a minimizer for a particular mass $M$, doubling the mass yields a minimizer with double the density. This is unbiological, and results from the attraction and the repulsion scaling equally with density.

An alternative aggregation model that does form well-spaced groups is
\begin{equation}
\label{eq:intro-tbl}
\dot{u} + \nabla \cdot (u \vec{v}) = 0, \quad \vec{v} = -\nabla Q_a * u - u \nabla u,
\end{equation}
initially proposed in \cite{TopBerLew2006}. While (\ref{eq:nonlocal-intro}) models attraction and repulsion nonlocally, (\ref{eq:intro-tbl}) models attraction nonlocally via a kernel $Q_a$, but models short-range repulsion via the (local) nonlinear diffusive velocity $-u \nabla u$. This model is similar to (\ref{eq:kawasaki}) but incorporates diffusion that is density-dependent and degenerate. That is, the diffusive flux tends to zero as the density tends to zero, which allows for compactly supported solutions.

Eq. (\ref{eq:intro-tbl}), like (\ref{eq:nonlocal-intro}), allows the formation of compactly supported solutions \cite{TopBerLew2006,Bed2011,BurFetHua2014}. It minimizes the energy,
\begin{equation}
\label{eq:energy-tbl-intro}
E(u) = \frac 12 \int \int u(\vec{x}) Q_a(|\vec{x}-\vec{y}|) u(\vec{y}) \, d\vec{y} \,d\vec{x} + \frac{1}{6} \int u^3  \, d\vec{x}.
\end{equation}
The nonlinear diffusion has two notable effects. First, because the repulsion dominates at higher densities, the density is bounded as the mass $M$ increases, leading to aggregations of approximately constant density at large mass~\cite{TopBerLew2006,BurFetHua2014}. Second, the introduction of the gradient as part of the repulsion in the social velocity smooths the density profile at the support's boundary. Solutions are continuous with a discountinuous gradient, often referred to as a non-zero contact angle. Further work on (\ref{eq:intro-tbl}), and on a generalization incorporating repulsion of the form $-u^p \nabla u$, has examined existence and uniqueness \cite{BerSle2010,BedRodBer2011,BedRod2014}, regularity \cite{ChaKimYao2013}, global attractors \cite{Gal2013}, and properties of steady states as a function of the power $p$ of the nonlinear repulsion \cite{BurFetHua2014}.

Despite the promise of (\ref{eq:intro-tbl}) as a model, it poses significant challenges. Beyond the analytical difficulties associated with the nonlocal operator, even numerical simulation is problematic. First, for 2-d simulation on an $n \times n$ grid, the convolution requires $\order(n^4)$ operations for quadrature or $\order(n^2 \log n)$ for pseudospectral evaluation. Second, the steady states of (\ref{eq:intro-tbl}) can have steep edges, like some steady states of (\ref{eq:nonlocal-intro}). These require a high degree of spatial resolution. Third, standard methods produce oscillations emanating from the contact points, and such artifacts must be eliminated. Fourth, and relatedly, a numerical scheme must preserve the positivity of the model. It is notable but perhaps not surprising that fully two dimensional simulations of (\ref{eq:intro-tbl}) are rare in the literature \cite{TopBerLew2006,CarCheHua2015}.  At present, most published simulations are either in 1-d or in a radially symmetric 2-d geometry. This is true of (\ref{eq:nonlocal-intro}) as well, excepting special cases where the computation reduces to a boundary integral problem describing the motion of a self-deforming curve \cite{TopBer2004,BerLauLeg2012}. Given the attractive features of (\ref{eq:intro-tbl}) from a modeling perspective, and yet given the analytical and computational challenges, we investigate whether there exists a simpler, local model that nonetheless preserves important features of (\ref{eq:intro-tbl}).

In summary of this discussion, (\ref{eq:intro-tbl}) produces groups whose density is bounded as mass $M$ increases but the model is difficult to analyze and simulate. There is a long mathematical history of approximating dynamics in the limit of slow variation to simplify governing equations \cite{CroHoh1993,OroDavBan1997}.  As we will describe in detail in this paper, an expansion of the convolution in (\ref{eq:intro-tbl}) the long-wave limit yields, after non-dimensionalization,
\begin{equation}
\label{eq:approximating}
Q_a * u \approx -  u - \nabla^2 u.
\end{equation}
Retaining only the first term in this expansion yields a nonlinear diffusion equation
\begin{equation}
\label{eq:intro-bd}
\dot{u} + \nabla \cdot (u \vec{v}) = 0, \quad \vec{v} =   (1-u) \nabla u,
\end{equation}
which is ill-posed due to negative (backward) diffusion at small densities. However, retaining the second term in the expansion yields stabilizing fourth-order degenerate diffusion.  We will concentrate on this degenerate fourth-order equation,
\begin{equation}
\label{eq:intro-dch}
\dot{u} + \nabla \cdot (u \vec{v}) = 0, \quad \vec{v} = - u \nabla u + \nabla (  u +  \nabla^2 u).
\end{equation}
It is curious that for this model, well-posedness at short wavelengths comes from the fourth-order term derived from nonlocal attraction. As attraction operates at longer wavelengths than repulsion, one must retain higher derivatives to model it faithfully. Mathematically, the fourth-order term describes attractive forces that are responsive not only to variations in density but to the convexity of the density. This yields a social force analogous to a surface tension, and has the effect of damping variation in the density. A moment expansion similar to the approximation (\ref{eq:approximating}) is used to derive a different model for the movement of ecological populations in \cite{Lew1994}.

A great deal is known about (\ref{eq:intro-dch}). As we will discuss below, it is related to both the Cahn-Hilliard equation \cite{Nov2008} which arises as a model of phase separation and to thin-film equations which model surface-tension driven wetting of a substrate by a viscous fluid \cite{OroDavBan1997,Mye1998,CraMat2009}. While the existence theory for our model is more strongly related to previous work on thin film equations, the dynamics are akin to phase separation and therefore we refer to our model as a Cahn-Hilliard model. Our model (\ref{eq:intro-dch}) also may be considered a degenerate Sivashinsky equation \cite{Siv1983,NovGri1995}.

After nondimensionalization, (\ref{eq:intro-dch}) is equivalent to
\begin{equation}
\label{eq:intro-dch1}
\dot{u} =  \nabla \cdot \left( u \nabla \left[ \frac{u^2}{2}-u -\nabla^2 u  \right] \right),
\end{equation}
which reveals a little more of its structure. It is the composition of a negative-definite second-order self-adjoint operator ($ \nabla \cdot  u \nabla$) with the first variation of an energy,
\begin{equation}
\label{eq:energy-dch-intro}
E(u) = \int  \frac 12 |\nabla u|^2 -\frac{u^2}{2} +\frac{u^3}{6}  \, d\vec{x}.
\end{equation}
This equation is an example of the more general class of Cahn-Hilliard models \cite{Nov2008} which take the form,
\begin{equation}
\label{eq:ch}
\dot{u} = \nabla \cdot \left( M(u) \nabla \left[f(u) - \nabla^2 u\right] \right).
\end{equation}
Here, $M(u)$ is a non-negative mobility, so $\nabla \cdot M(u) \nabla$ is a negative-definite second-order self-adjoint operator. The equation minimizes
an energy
\begin{equation}
\label{eq:intro-ch-energy}
E(u) = \int  \frac 12 |\nabla u|^2 +F(u)  \, d\vec{x},
\end{equation}
where $F(u)= \int f(u)\,du$ is most commonly taken to be a double well potential in the Cahn-Hilliard literature. Eq. (\ref{eq:ch}) was first developed in \cite{CahHil1958,Cah1959}  to describe binary alloys. A hallmark of this model is that it exhibits phase separation, where the density segregates into regions in which $u$ assumes one of the two minimum values of $F$, and these regions are separated by narrow transition layers \cite{Cah1961, Cah1965, Gra1993}.

Our model (\ref{eq:intro-dch1}) borrows a feature more familiar in the thin film context. Biologically, the density $u$ must remain nonnegative, much like a fluid film thickness. In our model, this constraint is enforced by the degenerate mobility $M(u)\equiv u$. This feature not only ensures that the density remains nonnegative, but it also allows regions of zero density, akin to dry spots in fluid films. Degenerate mobilities have been considered in both the Cahn-Hilliard context \cite{EllGar1996,Nov2008} and in the thin films context \cite{BerFri1990,HocRos1993,BerBreDup1994,BerBerDal1995,BerPug1996,BerPug1998,LauPug2000,Gla2003,BerJuLu2011}. 

The work in \cite{NovShi2010} nicely summarizes and extends the analytical progress that has been made over the past two decades on a class of thin film equations including (\ref{eq:intro-dch1}), primarily in one spatial dimension. For our model in one dimension, \cite{NovShi2010} shows that there exist nonnegative solutions with compact and potentially disjoint support. Where $u$ is positive, these solutions are infinitely differentiable. Moreover, these solutions are continuously differentiable everywhere. Stated differently, the slope at the contact point, where $u$ touches down to zero, is zero. Additionally, the touchdown is generically quadratic and the speed of propagation of the contact points is finite. To our knowledge, theory for uniqueness and existence in higher dimensions is lacking. Still, motivated by these one dimensional results, we will primarily consider solutions that are infinitely differentiable for positive values and touch down with zero contact slope.

The rest of this paper is organized as follows. In Section \ref{sec:derivation}, we perform a long-wave expansion of (\ref{eq:intro-tbl}) leading to the degenerate Cahn-Hilliard equation (\ref{eq:intro-dch}), augmenting both models with an external potential describing the environment. Section \ref{sec:characteristics} calculates the linear stability of the two models, builds a framework for examining them as energy minimization problems, and describes how numerical computation of minimizers will be performed. Section \ref{sec:minimizers} studies in detail the minimizers for the case of no external potential. In brief, minimizers of both models share many properties, including compact support and a common peak density in the large mass limit. With the Cahn-Hilliard model established as a reasonable approximation of the nonlocal one, in Section \ref{sec:gaussian} we consider an external potential and show that when a potential well is sufficiently steep, it can be replaced with an \emph{approximate obstacle potential} consisting of regions where the potential is zero separated by inhibited regions where the potential is effectively infinite. Finally, Section \ref{sec:randper} uses these obstacle potentials to investigate how peak population density is affected by resource distribution. A random distribution of resources tends to increase peak density above its intrinsic value while a periodic distribution can decrease it.

\section{Derivation of the local model}
\label{sec:derivation}

We begin with the nonlocal model of \cite{TopBerLew2006} augmented with an external potential $W(\vec{x})$ which models taxis due to environmental factors,
\begin{equation}
\label{eq:tbldimensioned}
\dot{u} + \nabla \cdot (u \vec{v}) = 0, \quad \vec{v} = -\nabla Q_a * u - Ru \nabla u -\nabla W,
\end{equation}
where $u(\vec{x},t)$ is the population density and $\vec{x} \in \mathbb{R}^n$. The parameter $R>0$ controls the strength of short-range repulsion. For now, we assume the potential $W$ to be continuously differentiable, although eventually we will consider a discontinuous limit. The kernel $Q_a$ in the convolution above describes social attraction over finite (rather than infinitesimal) distances. We place several assumptions on this kernel:
\begin{enumerate}
\item $Q_a$ is radial, that is, $Q_a \equiv Q_a(r)$, $r=|\vec{x}|$. This is the case for social interactions that are isotropic in space.
\item $Q_a'(r) \geq 0$ so that $Q_a$ describes (pure) attraction.
\item As $|r| \to \infty$, $Q_a$ approaches a constant since organisms at very long distances do not interact with each other. Without loss of generality, we assume $Q_a \to 0$ as $|r| \to \infty$.
\item $Q_a \leq 0$ as a result of the previous two assumptions.
\end{enumerate}
This is a dimensioned model; by rescaling $\vec{x}$, $t$, and $u$, we may write a dimensionless version of (\ref{eq:tbldimensioned}) as
\begin{equation}
\label{eq:tblswarm}
\dot{u} + \nabla \cdot (u \vec{v}) = 0, \quad \vec{v} = -\nabla Q_a * u - u \nabla u -\nabla W,
\end{equation}
where we have chosen $R=1$. Also, we choose our nondimensionalization such that $Q_a$ has unit volume, that is,
\begin{equation}
M_0 = \int_{\mathbb{R}^n} Q_a(r)\,d\vec{x} = -1,
\end{equation}
where $M_0$ denotes the zeroth moment. In addition, we specify the second moment of the potential in order that $Q_a$ has a length scale of order unity. This condition is
\begin{equation}
\label{eq:Qscaling}
M_2 = \int_{\mathbb{R}^n} Q_a (r)\,r^2\,d\vec{x} = -2n.
\end{equation}
This rescaling, which we take without loss of generality, will be convenient for our subsequent calculations.

Presently, we will approximate (\ref{eq:tblswarm}) with a local partial differential equation by applying the convolution theorem to $-\nabla Q_a * u$ and performing a long-wave expansion in Fourier space \cite{Mur2002ch11,LevTopBer2009}. Define the Fourier transform of a function $f(\vec{x})$,
\begin{equation}
\mathfrak{F}(f) = \widehat{f} = \int_{\mathbb{R}^n} f(\vec{x}) \exp{-i \vec{k} \cdot \vec{x}}\,d\vec{x}.
\end{equation}
Since our kernel $Q_a$ is radial in physical space, its Fourier transform is radial in the transformed variable $\vec{k}$. So
\begin{equation}
\widehat{Q}_a(k) = \int_{\mathbb{R}^n} Q_a(r) \exp{-i \vec{k} \cdot \vec{x}}\,d\vec{x}, \label{eq:Qahat}
\end{equation}
where $k=|\vec{k}|$ and $r = |\vec{x}|$. Expanding the complex exponential,
\begin{equation}
\exp{-i \vec{k}\cdot\vec{x}} = \sum_{m=0}^{\infty} (-i)^m \frac{(\vec{k}\cdot\vec{x})^m}{m!} = 1 - i\vec{k}\cdot\vec{x} - \frac{1}{2}(\vec{k}\cdot\vec{x})^2 + \cdots,
\end{equation}
and substituting into (\ref{eq:Qahat}) will yield a moment expansion for $\widehat{Q}_a$. We will retain only the first three terms of this expansion. For the first term,
\begin{equation}
\int_{\mathbb{R}^n} Q_a(r) \,d\vec{x} = M_0 = -1,
\end{equation}
due to our choice of scaling. For the second term,
\begin{equation}
-i \int_{\mathbb{R}^n} (\vec{k}\cdot\vec{x}) Q_a(r)\,d\vec{x} = 0,
\end{equation}
because the term linear in $\vec{k}$ vanishes due to symmetry; in fact, all terms odd in $\vec{k}$ vanish. Finally, for the third (quadratic in $\vec{k}$) term, we have
\begin{subequations}
\label{eq:quadratic}
\begin{eqnarray}
-\frac{1}{2} \int_{\mathbb{R}^n} (\vec{k}\cdot\vec{x})^2 Q_a(r)\,d\vec{x} & = & -\frac{1}{2} \sum_{p=1}^n \sum_{q=1}^n k_p k_q \left[ \int _{\mathbb{R}^n} Q_a(r) x_p x_q \,d\vec{x}\right],\label{eq:quadratic1} \\ 
& = & -\frac{1}{2} \sum_{p=1}^n k_p^2 \left[ \int_{\mathbb{R}^n} Q_a(r) x_p^2 \,d\vec{x}\right],\label{eq:quadratic2}\\
& = & -\frac{1}{2} \sum_{p=1}^n k_p^2 \left[ \frac{1}{n} \int_{\mathbb{R}^n} Q_a(r) r^2 \,d\vec{x}\right],\label{eq:quadratic3}\\
& = & -\frac{k^2}{2n} M_2, \label{eq:quadratic4}\\ 
& = & k^2.\label{eq:quadratic5}
\end{eqnarray}
\end{subequations}
Here, the right hand side of (\ref{eq:quadratic1}) expands the square of the dot product as a double sum, (\ref{eq:quadratic2}) utilizes symmetry to recognize the vanishing non-diagonal moments, (\ref{eq:quadratic3}) uses radial symmetry of $Q_a$, (\ref{eq:quadratic4}) substitutes the definition of $M_2$, and (\ref{eq:quadratic5}) uses the normalization of $M_2$ in (\ref{eq:Qscaling}).

Putting these results together,
\begin{equation}
\label{eq:long-wave}
\widehat{Q}_a = -1 + k^2 + \order(k^4),
\end{equation}
and therefore, we approximate
\begin{equation}
Q_a * u \approx -u - \nabla^2 u, \label{eq:long-wave2}
\end{equation}
which constitutes a long-wave approximation. With this approximation, the original governing equation (\ref{eq:tblswarm}) becomes the new, local equation
\begin{equation}
\dot{u} + \nabla \cdot (u \vec{v}) = 0, \quad \vec{v} = \nabla(u + \nabla^2 u -W) - u \nabla u.
\end{equation}
This equation may be rearranged as
\begin{equation}
\label{eq:ge}
\dot{u} = \nabla \cdot \left( u \nabla \left[\frac{1}{2}u^2 +W- u - \nabla^2 u\right] \right),
\end{equation}
which is a Cahn-Hilliard equation of type (\ref{eq:ch}) with degenerate mobility $M(u)=u$, $f(u) = u^2/2 - u$, and an external potential $W(\vec{x})$. We will explore the degree to which minimizers of this truncated model approximate those of the nonlocal model (\ref{eq:tblswarm}). While the former are easier to compute owing to the purely local interaction rules, it is important to realize what is lost in the local approximation. In (\ref{eq:tblswarm}), disjoint clumps of density interact with each other exponentially weakly owing to the nonlocality. However, in (\ref{eq:ge}), disjoint clumps do not interact at all owing to the local behavior and degenerate diffusion.

\section{Basic model characteristics}
\label{sec:characteristics}
To recapitulate, we are studying the nonlocal model
\begin{equation}
\label{eq:NL3}
\dot{u} = \nabla \cdot \left( u \nabla \left[Q_a * u + \frac{1}{2}u^2 +W \right] \right),
\end{equation}
and the degenerate Cahn-Hilliard equation which is a long-wave approximation to it,
\begin{equation}
\label{eq:CH3}
\dot{u} = \nabla \cdot \left( u \nabla \left[- u - \nabla^2 u + \frac{1}{2}u^2 +W \right] \right),
\end{equation}
where $u(\vec{x},t)$ is a non-negative density.
To complete the formulation of the problem, one must specify the domain, the boundary conditions, the external potential $W$ and initial conditions for the density $u$. For much of this paper we consider all of $\mathbb{R}^n$ or periodic domains in one and two dimensions. Solutions to problems with degenerate diffusion often have compact support \cite{ZelBar1958,Aro1986,Gla2003}. For the Cahn-Hilliard problem (\ref{eq:CH3}) we expect $u(x,t)$ to be continuously differentiable everywhere and infinitely differentiable for all $\vec{x}$ in the support of $u$. For the nonlocal model (\ref{eq:NL3}), we expect $u(x,t)$ may have jump discontinuities at the edge of its support, but will be infinitely differentiable for all $\vec{x}$ in the support of $u$ \cite{BerTop2011,BerTop2013}. Leveraging the compact support of the solution, when we wish to consider densities confined to a finite domain rather than specifying boundary conditions at the edge of a domain, we will either consider a large periodic domain or we will specify that $W(\vec{x})$ is sufficiently large to drive the density to zero except in some subdomain of interest.

Both models are conservation laws which are well known to conserve total mass (see, \emph{e.g.}, \cite{TopBerLew2006}). If
\begin{equation}
M = \int_\Omega u ~ d \vec{x},
\end{equation}
where $\Omega$ is the domain (or the support of the solution) then 
\begin{equation}
\frac{dM}{dt} = 0.
\end{equation}

For many problems with a reflection symmetry, including our models (\ref{eq:NL3}) and (\ref{eq:CH3}), the location of the center of mass will also be conserved when the domain is $\mathbb{R}^n$ and the mass is finite. In this case, for  both (\ref{eq:NL3}) and (\ref{eq:CH3}), if we define
\begin{equation}
\vec{X}  = \frac{1}{M} \int_\Omega \vec{x} u ~ d \vec{x},
\end{equation}
a straightforward calculation shows that 
\begin{equation}
\frac{d\vec{X}}{dt} = \frac{1}{M} \int_\Omega \vec{x} \dot{u} ~ d \vec{x} = - \frac{1}{M} \int_\Omega u \nabla W ~ d \vec{x},
\end{equation}
from which we deduce that in the absence of an external potential, the center of mass will be stationary. In the presence of an external potential, mass tends to migrate towards minima of the potential.

\subsection{Linear stability} In the absence of an external potential ($W=0$), both (\ref{eq:NL3}) and (\ref{eq:CH3}) admit steady states with constant density $\bar{u} \geq 0$. To analyze their  linear stability, let $u(\vec{x},t)=\bar{u} +\tilde{u}(x,t)$, linearize the equations in $\tilde{u}$, and let $\tilde{u} \propto \exp{i\vec{k}\cdot\vec{x}+\sigma (k) t}$ ($k = |\vec{k}|$) to yield the dispersion relation $\sigma(k)$. For both models, $\sigma(0) =  0$ for all $\bar{u}$ because the models are of conservation form \cite{CroHoh1993}.

For the nonlocal model (\ref{eq:NL3}),
\begin{equation}
\label{eq:tbldisprel}
\sigma(k) = -k^2 \bar{u} \left[\bar{u} - \widehat{Q}_a\right],
\end{equation}
a result also stated in \cite{TopBerLew2006}. The Fourier transform of the attractive kernel $Q_a$ depends on spatial dimension. From our nondimensionalization in Section \ref{sec:derivation},
\begin{equation}
\widehat{Q}_a = -1 +k^2 +\order(k^4).
\end{equation}
This guarantees that the dispersion relations for (\ref{eq:NL3}) and (\ref{eq:CH3}) are identical to $\order(k^4)$.

To further analyze the model one must specify $Q_a$. A commonly used example which we also adopt is the Laplace potential.  In our nondimensionalization, this potential is
\begin{equation}
\label{eq:laplacepotential}
Q_a(\vec{x}) = - c_1 \exp{-c_2 |\vec{x}|},
\end{equation}
where
\begin{equation}
c_1 = \frac{\sqrt{\pi}}{\Gamma \left( \frac{n+1}{2}\right)} \left( \frac{n+1}{8\pi} \right)^{n/2}, \quad c_2 = \sqrt{\frac{n+1}{2}},
\end{equation}
in which case
\begin{equation}
\widehat{Q}_a(k) = - \left(1+ \frac{2k^2}{n+1} \right) ^ {-\frac{n+1}{2}},
\end{equation}
which is a power of the Lorentzian function. We see that $\widehat{Q}_a(0)=-1$ and $\widehat{Q}_a(k)$ is increasing in $k$,  and tends to zero. Consequently, if $\bar{u} > 1$, $\sigma(k) < 0$ for all $k > 0$ and the homogeneous steady state is linearly stable. However, if $\bar{u} < 1$, there is a long-wave instability. There exists a finite, continuous band of wave numbers extending from $k = 0$ to
\begin{equation}
k = \sqrt{\frac{n+1}{2}\left(\bar{u}^{-2/(n+1)}-1  \right)},
\end{equation}
for which $\sigma(k) >0$.

For the Cahn-Hilliard model (\ref{eq:CH3}), the calculations are significantly simpler. We obtain
\begin{equation}
\label{eq:disprel}
\sigma(k) = -k^2 \bar{u} \left[(\bar{u}-1)+k^2\right].
\end{equation}
The stability theory is analogous to the nonlinear model. If $\bar{u} > 1$, $\sigma(k) < 0$ for all $k > 0$ and the homogeneous steady state is linearly stable. However, if $\bar{u} < 1$, there is a long-wave instability. There exists a finite, continuous band of wave numbers extending from $k = 0$ to $k = \sqrt{1- \bar{u}}$ for which $\sigma(k) >0$. 

It is worthwhile to consider why a fourth-order truncation of (\ref{eq:NL3}) is superior to a second-order one. For a second-order truncation, the dispersion relation is simply
\begin{equation}
\label{eq:disprel2}
\sigma(k) = -k^2 \bar{u}(\bar{u}-1).
\end{equation}
The stability threshold is still $\bar{u} = 1$ as above. However, in the linearly unstable regime, \emph{all} modes have positive growth rates which grow unboundedly with $k$, similar to the backwards heat equation. Thus, the linear problem for the second-order truncation is ill-posed in this case, shedding light on why an approximation to fourth order is desirable. Our fourth-order truncation is analogous to the strategy used in deriving amplitude equations for marginally long-wave unstable systems via modulation theory \cite{CroHoh1993}. Retaining the destabilizing second-order and stabilizing fourth-order terms yields the most parsimonious truncation that is linearly well-posed.

\subsection{Energy} We will show that both the nonlocal model (\ref{eq:NL3}) and the Cahn-Hillard model (\ref{eq:CH3}) can be interpreted as gradient flows. Both may be written in the form
\begin{equation}
\label{eq:Eform}
\dot{u} = \nabla \cdot u  \nabla \left[\Lop u +f(u) + W \right],
\end{equation}
where $\Lop$ is a linear operator. For (\ref{eq:NL3}), $\Lop u = Q_a * u$ and for (\ref{eq:CH3}), $\Lop u = -\nabla^2 u - u$. For both models, $f(u) = u^2/2$. In (\ref{eq:Eform}), $\nabla \cdot u  \nabla$ is a negative-definite self-adjoint second-order operator on non-negative functions $u$.

We now define quantities useful for constructing an energy. First, define a symmetric quadratic form for which $\Lop u$ is the first variation. For (\ref{eq:NL3}) define
\begin{equation}
\Qop[u,v] \equiv \frac{1}{2}  \int_\Omega u Q_a*v\,d\vec{x} = \frac{1}{2} \int_\Omega \int_\Omega Q_a(|\vec{x}-\vec{y}|)u(\vec{x})v(\vec{y})\,d\vec{x}\,d\vec{y},
\end{equation}
and for (\ref{eq:CH3}) define
\begin{equation}
\Qop[u,v] \equiv \frac{1}{2} \int_\Omega \nabla u \cdot \nabla v - uv\,d\vec{x}.
\end{equation}
Next, choose
\begin{equation}
F(u) \equiv \int_\Omega u^3/6\,d\vec{x},
\end{equation}
whose first variation is $f(u)$.

The total energy for both models is
\begin{equation}
\label{eq:energy}
E(u) = \Qop[u,u] + F(u) + \int_\Omega Wu\,d\vec{x}.
\end{equation}
The first variation of the energy is
\begin{equation}
\label{eq:firstvariation}
\frac{\delta E}{\delta u} = \Lop u + f(u) + W,
\end{equation}
and both evolution equations can be written as
\begin{equation}
u_t = \nabla \cdot u \nabla \left[ \frac{\delta E}{\delta u}\right].
\end{equation}
It follows that the evolution is a gradient flow in a weighted metric \cite{Gla2003,Vil2003,BerTop2011,BerTop2013,CarChiHua2014,SimSleTop2015}. The time derivative of the energy is
\begin{subequations}
\label{eq:energygradient}
\begin{eqnarray}
\frac{dE}{dt}  &=& \phantom{-}\int_\Omega \frac{\delta E}{\delta u} u_t\,d\vec{x},\\
&=&  \phantom{-} \int_\Omega \frac{\delta E}{\delta u}  \nabla \cdot u  \nabla \frac{\delta E}{\delta u}\,d\vec{x},\\
&=&  - \int_\Omega u \left| \nabla \frac{\delta E}{\delta u} \right|^2\,d\vec{x}.
\end{eqnarray}
\end{subequations}
Consequently, energy is decreasing except for stationary states where $\delta E/\delta u$ is constant on the support of the solution.

\subsection{Energy minimization and numerics}
Because our governing equations are gradient flows, we expect almost all initial conditions to evolve to energy minimizers. We exploit energy minimization to explore the equilibria of both models, namely (\ref{eq:NL3}) with the Laplace potential (\ref{eq:laplacepotential}), and (\ref{eq:CH3}). This strategy plays a prominent role for the remainder of this paper. In order for a solution $u$ to be an equilibrium, it must minimize the energy (\ref{eq:energy}), satisfy the appropriate mass constraint, and be non-negative. In summary, the constrained minimization problem is
\begin{subequations}
\label{eq:minconds}
\begin{gather}
\min_{u(\vec{x})} E(u) \equiv \Qop[u,u] + F(u)  + \int_\Omega Wu\,d\vec{x},\\
\intertext{subject to the constraints:}
 \int_\Omega u\,d\vec{x} = M, \\
u(\vec{x}) \geq 0 \text{ for all $x \in \Omega$}.
\end{gather}
\end{subequations}

We solve this problem using the \texttt{fmincon} minimization routine of \texttt{M\sc{atlab}}'s optimization toolbox, which accepts an objective function as well as constraints. We use periodic domains in one and two dimensions with equally spaced grid points. For the nonlocal model on a periodic domain, we replace the attractive potential $Q_a$ with its periodized analog and evaluate the convolution in $\Qop$ via matrix multiplication. For the Cahn-Hilliard model, we evaluate derivatives using a centered difference on each interval in one dimension, or each cell in two dimensions. To aid convergence of the algorithm, we also supply the minimization algorithm with the gradient of our collocated approximation of the energy. We typically choose a random initial condition to seed the algorithm. Sometimes we solve the minimization problem on a coarse grid and then refine the grid by factors of two in order to obtain highly-resolved final states.

Our methodology neglects the dynamics of the approach to equilibria, and relatedly, the issue of whether those equilibria are accessible on relevant biological time scales. Nonetheless, we see the exploration of minimizers as an important first step in understanding (\ref{eq:NL3}) and (\ref{eq:CH3}).

\section{Energy minimizers in the absence of an external potential}
\label{sec:minimizers}

\subsection{Motivation and numerical computations}

Recalling that the nonlocal equation (\ref{eq:NL3}) and the degenerate Cahn-Hilliard equation (\ref{eq:CH3}) have steady, compactly supported solutions and motivated by the aggregation of biological organisms, we now study the simplest case of a single clump having total mass $M$ on an effectively infinite domain with no external potential ($W=0$). A phase plane analysis of (\ref{eq:NL3}) suggests that in one spatial dimension, for each mass $M$, there is a single energy-minimizing clump \cite{TopBerLew2006}. For a generalization of (\ref{eq:NL3}) to diffusion of power-law form, rigorous proofs show existence and uniqueness in higher dimensions, subject to appropriate assumptions \cite{Bed2011,BedRodBer2011}. Similarly, degenerate Cahn-Hilliard models such as (\ref{eq:CH3}) are gradient flows that minimize an energy; see \cite{LauPug2000, Nov2008} and references therein. Our primary strategy for studying (\ref{eq:NL3}) and (\ref{eq:CH3}) will be to exploit energy minimization.

To begin, we numerically compute minimizers as described in Section \ref{sec:characteristics}, varying total mass $M$ as our parameter for both models, and in both one and two dimensions. When varying $M$, we choose values of domain length $L$ such that the spatially homogeneous solution is linearly unstable to a single mode. Strictly speaking, the minimizers of the nonlocal model (\ref{eq:NL3}) are affected by their periodic images in our computation. In practice, with our choice of attractive potential (\ref{eq:laplacepotential}), these effects are exponentially small and invisible on the plots shown. For the Cahn-Hilliard model (\ref{eq:CH3}), the compactly supported solution does not interact with its periodic images.

\begin{figure}[t]
\begin{tikzpicture}
\begin{semilogxaxis}[xlabel=$M$,ylabel=$\|u\|_\infty$,y label style={rotate=-0},width = \textwidth,height = 0.45\textwidth,xmin=0.1,xmax=100,ymin=0,ymax=2,enlargelimits=0]
\pgfplotstableread{data/varymass1d.txt}\massdata;
\addplot[mark=o,solid] table[x=M,y=umaxch] {\massdata};
\addplot[mark=square,dashed] table[x=M,y=umaxnl] {\massdata};
\node at (rel axis cs:0.05,0.92) {$(a)$};
\end{semilogxaxis}
\end{tikzpicture}
\begin{tikzpicture}
\begin{axis}[xmode = log, ymode = log, xlabel=$M$,ylabel=$\|\supp(u)\|$,y label style={rotate=-0},width = \textwidth,height = 0.45\textwidth,xmin=0.1,xmax=100,ymin=0,ymax=100,enlargelimits=0]
\pgfplotstableread{data/varymass1d.txt}\massdata;
\addplot[mark=o,solid] table[x=M,y=usuppch] {\massdata};
\addplot[mark=square,dashed] table[x=M,y=usuppnl] {\massdata};
\node at (rel axis cs:0.05,0.92) {$(b)$};
\end{axis}
\end{tikzpicture}
\caption{\label{fig:varymass1d} Properties of energy minimizers of the nonlocal model (\ref{eq:NL3}) (shown as squares) and the Cahn-Hilliard model (\ref{eq:CH3}) (shown as circles). For each mass $M$, and for both models, we calculate a minimizer on a periodic domain of length $L$ (increasing with $M$ so as to be effectively infinite) using $10L$ grid points. The dashed and solid curves are to guide the eye. The two models agree well for large $M$. (a) Peak density $\|u\|_\infty$ increases with $M$ and saturates to a constant value of 1.5. (b)~The size of the support, $\|\supp(u)\|$, increases and grows linearly at large $M$.}
\end{figure}
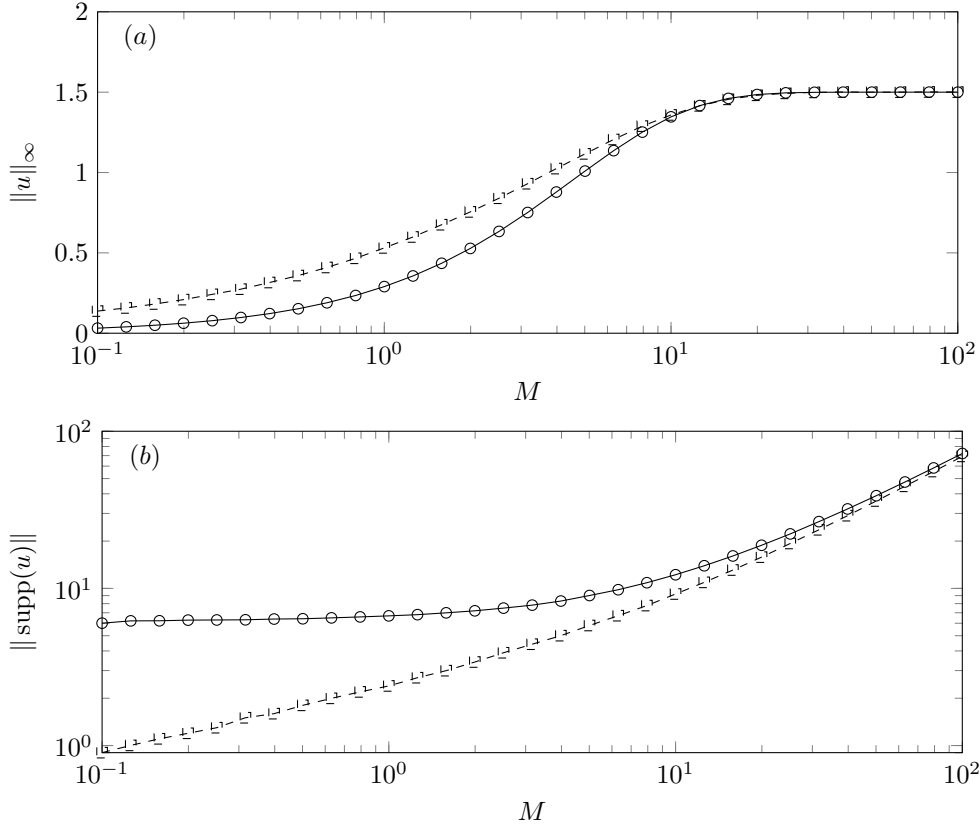

For both models in one dimension, example minimizers appear in Figure~\ref{fig:1dclumps}. Panel~(a) compares small mass solutions and (b) compares large mass solutions. The small $M$ solutions have well defined peaks, whereas the large $M$ solutions acquire a plateau-like structure; they are all compactly supported. The computational results for (\ref{eq:NL3}) and (\ref{eq:CH3}) differ at small masses, but coincide for large $M$, except in the transition region. Figure~\ref{fig:varymass1d} scans two key properties of the clumps as a function of $M$, namely the peak density $\|u\|_\infty$ and the size of the support $\|\supp(u)\|$. The peak density increases but saturates to a value of $1.5$ with increasing $M$; we will later analyze this important property. The support increases with $M$, and the growth appears to be linear at large $M$ (as we later confirm).

%%%%%%%%%%%%%%%%%%%%%%%%%%%%%%%%%%%%%%
\begin{figure}[t]
\begin{tikzpicture}
\begin{axis}[xlabel=$x$,ylabel=$u$,y label style={rotate=-90},width=0.48 \textwidth,height=0.48\textwidth,xmin=0,xmax=40,ymin=0,ymax=1.5,enlargelimits=0.02]
\pgfplotstableread{data/small1dclump.txt}\smallclump;
\addplot[no marks,solid] table[x=x,y=uch] {\smallclump};
\addplot[no marks,dashed] table[x=x,y=unl] {\smallclump};
\node at (rel axis cs:0.05,0.92) {$(a)$};
\end{axis}
\end{tikzpicture}
\hfill
\begin{tikzpicture}
\begin{axis}[xlabel=$x$,ylabel=$u$,y label style={rotate=-90},width=0.48 \textwidth,height=0.48\textwidth,xmin=0,xmax=40,ymin=0,ymax=1.5,enlargelimits=0.02]
\pgfplotstableread{data/large1dclump.txt}\bigclump;
\addplot[no marks,solid] table[x=x,y=uch] {\bigclump};
\addplot[no marks,dashed] table[x=x,y=unl] {\bigclump};
\node at (rel axis cs:0.05,0.92) {$(b)$};
\end{axis}
\end{tikzpicture}
\caption{\label{fig:1dclumps} Numerically calculated energy minimizers in one dimension. The nonlocal model result for (\ref{eq:NL3}) appears as a dashed curve, and the Cahn-Hilliard model result for (\ref{eq:CH3}) appears as a solid curve. (a) Mass $M = 4$, periodic domain length $L = 40$, calculated with $N = 400$ grid points. The minimizers have a well-defined peak. (b) $M = 35$, $L = 40$, $N = 400$. The minimizers have a plateau of peak density $\|u\|_\infty \approx 1.5$, and the two models coincide well except in the transition region.}
\end{figure}
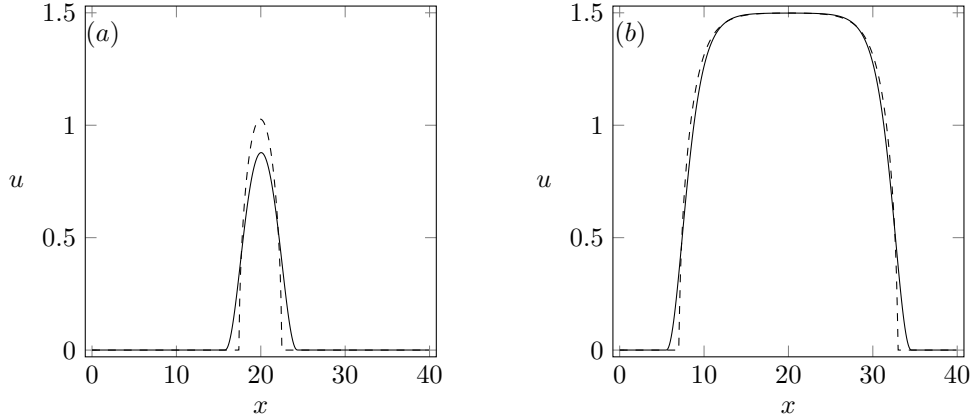
%%%%%%%%%%%%%%%%%%%%%%%%%%%%%%%%%%%%%%

%%%%%%%%%%%%%%%%%%%%%%%%%%%%%%%%%%%%%%
%%%%%%%%%%%%%%%%%%%%%%%%%%%%%%%%%%%%%%

\begin{figure}[t]
\begin{tikzpicture}
\begin{axis}[xlabel=$x$,ylabel=$y$,y label style={rotate=-90},width = 0.4 \textwidth,height = 0.4\textwidth,xmin=0,xmax=40,ymin=0,ymax=40, colorbar, colormap/jet, point meta min = 0, point meta max =2]
\addplot graphics [xmin=0,xmax=40,ymin=0,ymax=40]{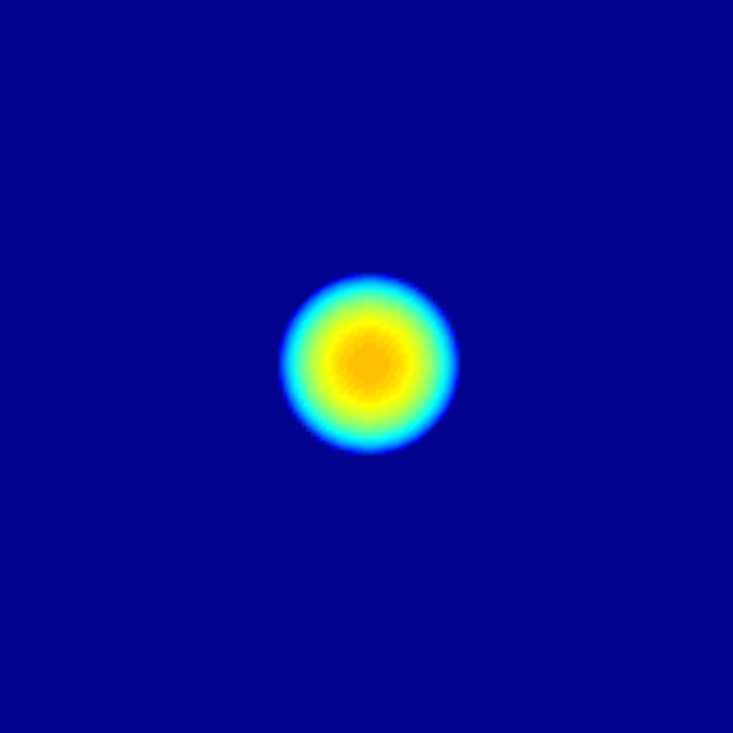};
\node at (rel axis cs:0.1,0.9) [color=white] {$(a)$};
\end{axis}
\end{tikzpicture}
\hfill
\begin{tikzpicture}
\begin{axis}[xlabel=$x$,ylabel=$y$,y label style={rotate=-90},width = 0.4 \textwidth,height = 0.4\textwidth,xmin=0,xmax=40,ymin=0,ymax=40, colorbar, colormap/jet, point meta min = 0, point meta max =2]
\addplot graphics [xmin=0,xmax=40,ymin=0,ymax=40]{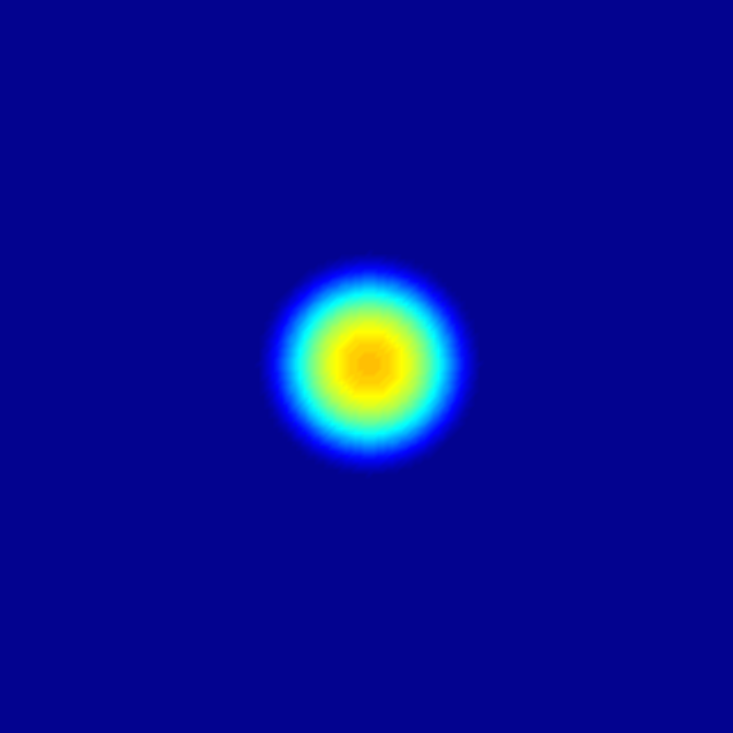};
\node at (rel axis cs:0.1,0.9) [color=white] {$(b)$};
\end{axis}
\end{tikzpicture}
\begin{tikzpicture}
\begin{axis}[xlabel=$x$,ylabel=$y$,y label style={rotate=-90},width = 0.4 \textwidth,height = 0.4\textwidth,xmin=0,xmax=40,ymin=0,ymax=40, colorbar, colormap/jet, point meta min = 0, point meta max =2]
\addplot graphics [xmin=0,xmax=40,ymin=0,ymax=40]{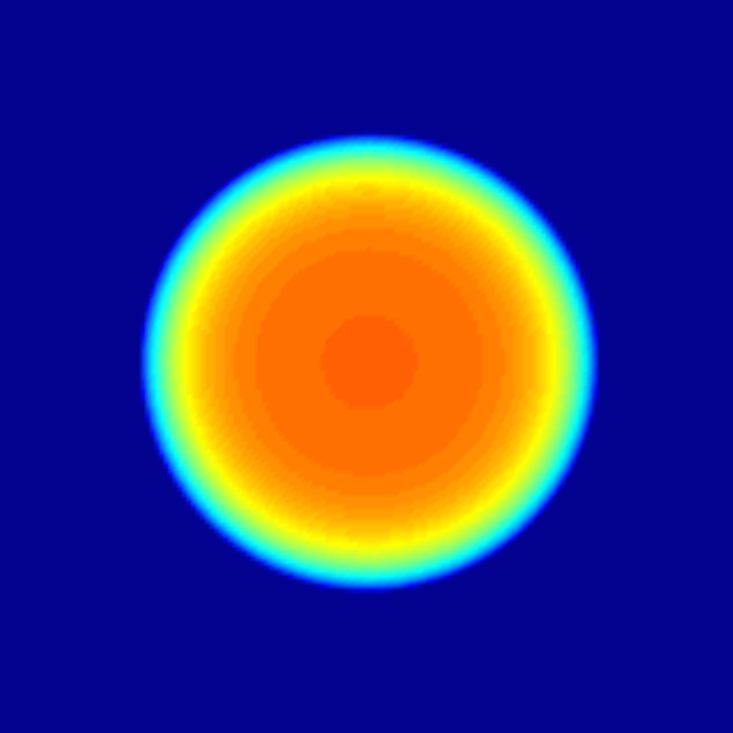};
\node at (rel axis cs:0.1,0.9) [color=white] {$(c)$};
\end{axis}
\end{tikzpicture}
\hfill
\begin{tikzpicture}
\begin{axis}[xlabel=$x$,ylabel=$y$,y label style={rotate=-90},width = 0.4 \textwidth,height = 0.4\textwidth,xmin=0,xmax=40,ymin=0,ymax=40, colorbar, colormap/jet, point meta min = 0, point meta max =2]
\addplot graphics [xmin=0,xmax=40,ymin=0,ymax=40]{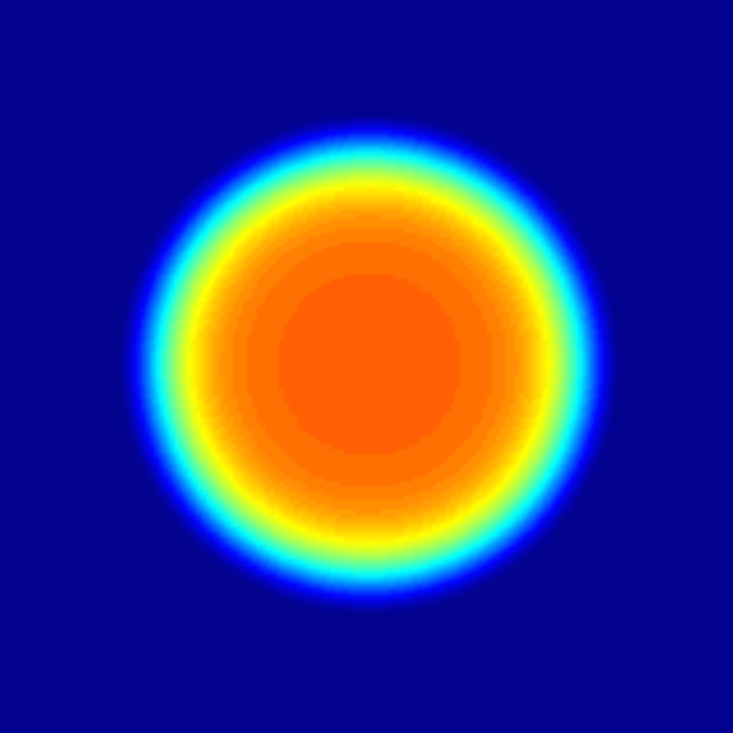};
\node at (rel axis cs:0.1,0.9) [color=white] {$(d)$};
\end{axis}
\end{tikzpicture}
\caption{\label{fig:2dclumps} Numerically calculated energy minimizers in two dimensions. Density $u$ is indicated by color. Minimizers of the nonlocal model (\ref{eq:NL3}) appear in (a) and (c), and minimizers of the Cahn-Hilliard model (\ref{eq:CH3}) appear in (b) and (d). (a,b) Mass $M = 70$, periodic box side length $L = 40$, calculated with $N = 160 \times 160$ grid points. The minimizers have a well-defined peak. (c,d) Like (a,b) but $M=600$.}
\end{figure}

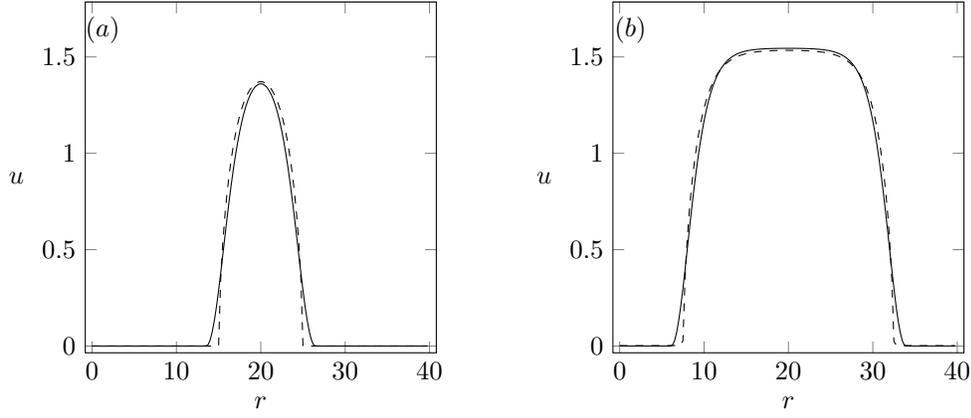
\begin{figure}[t]
\begin{tikzpicture}
\begin{axis}[xlabel=$r$,ylabel=$u$,y label style={rotate=-90},width = 0.48 \textwidth,height = 0.48\textwidth,xmin=0,xmax=40,ymin=0,ymax=1.75,enlargelimits=0.02]
\pgfplotstableread{data/smallradialclump.txt}\smallclump;
\addplot[no marks,solid] table[x=r,y=uch] {\smallclump};
\addplot[no marks,dashed] table[x=r,y=unl] {\smallclump};
\node at (rel axis cs:0.05,0.92) {$(a)$};
\end{axis}
\end{tikzpicture}
\hfill
\begin{tikzpicture}
\begin{axis}[xlabel=$r$,ylabel=$u$,y label style={rotate=-90},width = 0.48 \textwidth,height = 0.48\textwidth,xmin=0,xmax=40,ymin=0,ymax=1.75,enlargelimits=0.02]
\pgfplotstableread{data/largeradialclump.txt}\largeclump;
\addplot[no marks,solid] table[x=r,y=uch] {\largeclump};
\addplot[no marks,dashed] table[x=r,y=unl] {\largeclump};
\node at (rel axis cs:0.05,0.92) {$(b)$};
\end{axis}
\end{tikzpicture}
\caption{\label{fig:2dclumpsradial} Radial profiles of the radially-symmetric two dimensional minimizers in Figure~\ref{fig:2dclumps}. Minimizers of the nonlocal model (\ref{eq:NL3}) are dashed, and for the Cahn-Hilliard model (\ref{eq:CH3}), solid. (a)~Radial profiles for Figure~\ref{fig:2dclumps}(a,b). (b) Radial profiles for Figure~\ref{fig:2dclumps}(c,d). The plateau-like structure has peak density $\|u\|_\infty \approx 1.5$.}
\end{figure}

Example minimizers in a fully two dimensional geometry appear in Figure~\ref{fig:2dclumps}. The minimizers are radially symmetric. Panels (a) and (b) show clumps for small mass $M$ for (\ref{eq:NL3}) and (\ref{eq:CH3}) respectively. To give a better sense of the mass distribution, Figure~\ref{fig:2dclumpsradial}(a) shows the radial profiles for (\ref{eq:NL3}) (dashed) and (\ref{eq:CH3}) (solid). Figures~\ref{fig:2dclumps}(c) and (d) and Figure~\ref{fig:2dclumpsradial}(b) are analogous, but for large $M$. The results are similar to the one dimensional case. At small $M$, the clumps have a well-defined peak and at large $M$, they have a plateau of density $\|u\|_\infty \approx 1.5$. The two models agree well, especially for large $M$.

%%%%%%%%%%%%%%%%%%%%%%%%%%%%%%%%%%%%%%
%%%%%%%%%%%%%%%%%%%%%%%%%%%%%%%%%%%%%%

%%%%%%%%%%%%%%%%%%%%%%%%%%%%%%%%%%%%%%
%%%%%%%%%%%%%%%%%%%%%%%%%%%%%%%%%%%%%%

To better understand these results, it is helpful to analyze the steady state problem. To find conditions for energy minimizers, we recall the discussion of %, \emph{e.g.}, 
\cite{BerTop2011,BerTop2013}. Let
\begin{equation}
u(\vec{x}) = \bar{u} + \delta u.
\end{equation}
Here, $\bar{u}$ is an equilibrium solution of fixed mass $M$ and $\delta u$ is a small perturbation of zero mass, so that overall mass is conserved. Thus,
\begin{equation}
\int_\Omega \bar{u}\,dx = M,\quad \int_\Omega \delta u\,dx = 0,
\end{equation}
where $\Omega$ is the support of $\bar{u}$. Then to leading order, the difference in energy between the equilibrium solution and the perturbed one is
\begin{equation}
\label{eq:stationaryenergy}
\Delta E \equiv E(\bar{u} + \delta u) - E(\bar{u}) \approx \int_\Omega \frac{\delta E}{\delta u} \cdot \delta u\,dx.
\end{equation}
In order for the energy to be stationary, $\Delta E$ must vanish for all perturbations $\delta u$. As $\delta u$ has zero mass, a necessary and sufficient condition is that $\delta E/\delta u$ in (\ref{eq:stationaryenergy}) is a constant, which we call $\lambda$. That is,
\begin{equation}
\label{eq:critpoint}
\frac{\delta E}{\delta u} = \lambda,
\end{equation}
in the support of $u$. The quantity $\lambda$ has a physical interpretation: it is the energy per unit mass at each point in space \cite{BerTop2011,BerTop2013}. The condition (\ref{eq:critpoint}) is necessary for $u$ to be a minimizer. For sufficient conditions, one must examine the continuation of $\lambda$ outside of the support as well as the second variation of the energy. These ideas are discussed at length in \cite{BerTop2011, BerTop2013}. While we have presented numerical results thus far, for the remainder of this section, we analyze the minimizers of (\ref{eq:NL3}) and (\ref{eq:CH3}).

\subsection{Minimizers of the nonlocal equation in one dimension}

For (\ref{eq:NL3}), the energy (\ref{eq:energy}) in one dimension becomes
\begin{equation}
\label{eq:NLenergy}
E(u) = \int_\Omega \frac{1}{2}u[Q_a*u] + \frac{1}{6}u^3 + Wu\,dx.
\end{equation}
From (\ref{eq:firstvariation}) and (\ref{eq:critpoint}), and for $W=0$, minimizers satisfy
\begin{equation}
\label{eq:steadynonlocal}
\frac{\delta E}{\delta u} = \frac{1}{2}u^2 + Q_a * u = \lambda.
\end{equation}
To make analytical progress, we consider the small and large mass limits in turn, with some ideas following \cite{TopBerLew2006,BurFetHua2014}.

When $M$ is small, the characteristic length scale of the solution is small. Taylor expand the kernel as
\begin{equation}
Q_a \approx -\frac{1}{2} (1-|x|).
\end{equation}
Under this approximation, (\ref{eq:steadynonlocal}) becomes
\begin{equation}
\frac{1}{2}u^2 - \frac{M}{2} + \frac{1}{2} |x|*u = \lambda,
\end{equation}
where we have used the fact that $1*u = M$. Since $(|x|)_{xx} = 2\delta(x)$, differentiate twice to obtain
\begin{equation}
\left(\frac{u^2}{2}\right)_{xx} + u = 0.
\end{equation}
Without loss of generality, place the maximum of $u$ at the origin; this implies $u_x(0)=0$ by symmetry. For convenience, rescale $u$ by $u(0) = \|u\|_\infty$ and $x$ by $\|u\|_\infty^{1/2}$. Let $p = u/\|u\|_\infty$ and $\xi = x/\|u\|_\infty^{1/2}$, yielding
\begin{equation}
(p^2)_{\xi \xi} + 2p = 0,\quad p(0) = 1, \quad p_\xi(0) = 0.
\end{equation}
Multiply by $(p^2)_\xi$, integrate once, and apply the conditions at $\xi = 0$ to obtain the constant of integration, leading to
\begin{equation}
p_\xi^2 = \frac{2}{3} \left(\frac{1}{p^2} - p \right).
\end{equation}
Take the square root of both sides and separate variables to find an implicit solution,
\begin{equation}
\xi = \sqrt{\frac{3}{2}} \int_p^1 \frac{p\,dp}{\sqrt{1-p^3}}.
\end{equation}
As $p \to 0$, $\xi$ approaches the half-width in our rescaled variable, which implies
\begin{equation}
\label{eq:supp1}
\|\supp(u)\| = 2 \|u\|_\infty^{1/2} \left[ \sqrt{\frac{3}{2}} \int_0^1 \frac{p\,dp}{\sqrt{1-p^3}} \right] = \sqrt{\frac{6}{\pi}} \Gamma\left(\frac{5}{6}\right) \Gamma \left(\frac{2}{3}\right) \|u\|_\infty^{1/2}.
\end{equation}
The mass constraint is
\begin{equation}
\label{eq:mass1}
M = 2 \int_0^{\|\supp(u)\|/2} u(x)\,dx = \sqrt{6} \|u\|_\infty^{3/2} \int_0^1 \frac{p^2\,dp}{\sqrt{1-p^3}} = \sqrt{\frac{8}{3}}\|u\|_\infty^{3/2}.
\end{equation}
By combining (\ref{eq:supp1}) and (\ref{eq:mass1}), write $\|\supp(u)\|$ and $\|u\|_\infty$ as functions of mass $M$,
\begin{subequations}
\begin{gather}
\|u\|_\infty = \left( \frac{3}{8} \right)^{1/3} M^{2/3} \approx 0.721 M^{2/3},\\
\|\supp(u)\| = \sqrt{\frac{6}{\pi}} \Gamma\left(\frac{5}{6}\right) \Gamma \left(\frac{2}{3}\right) \left( \frac{3}{8} \right)^{1/6} M^{1/3} \approx 1.794 M^{1/3}.
\end{gather}
\end{subequations}

For large mass, recall from Figures~\ref{fig:1dclumps} and~\ref{fig:varymass1d} that solutions approach wide plateaus with peak density $\|u\|_\infty\approx1.5$. To understand this behavior, we analyze the large $M$ limit. When $M$ is large, spatial scales are large. Thus $Q_a \approx -\delta(\vec{x})$ and (\ref{eq:steadynonlocal}) becomes
\begin{equation}
\frac{\delta E}{\delta u} = \frac{1}{2}u^2 - u = \lambda,
\end{equation}
from which it immediately follows that $u$ is constant on the support of the solution. To determine what value that constant has, explicitly minimize the energy (\ref{eq:energy}) subject to the mass constraint. The energy is
\begin{equation}
\int_\Omega \frac{u^3}{6} - \frac{u^2}{2}\,d\vec{x} = \left( \frac{u^3}{6} - \frac{u^2}{2} \right) \|\supp(u)\|,
\end{equation}
However, the mass constraint dictates that for a this solution, $\|\supp(u)\| = M/u$, which we substitute to obtain
\begin{equation}
E = M\left(\frac{u^2}{6} - \frac{u}{2} \right).
\end{equation}
Thus
\begin{equation}
\frac{dE}{du} = M\left(\frac{u}{3} - \frac{1}{2} \right),
\end{equation}
which yields the critical point $u=3/2$. Since the solution is constant, $\|u\||_\infty = 3/2$, and $\|\supp(u)\| = 2M/3$. A curious feature of the approximation we have used is that it is invariant under any area-preserving map of the spatial domain; put differently, the approximation suggests that any arrangement of disjoint clumps having constant density $u=3/2$ has the same energy. However, we know there is an energetic cost for each clump (neglected in our analysis) proportional to the measure of its perimeter. Thus, we expect that the global minimizer is a rectangular profile of density $u = 3/2$.

\subsection{Minimizers of the Cahn-Hilliard equation in one dimension}
\label{sec:ch1d}

For the Cahn-Hilliard model (\ref{eq:CH3}), the energy (\ref{eq:energy}) in one dimension becomes
\begin{equation}
\label{eq:CHenergy}
E(u) = \int_\Omega \frac{1}{2}u_x^2 - \frac{1}{2}u^2 + \frac{1}{6}u^3 + Wu\,dx.
\end{equation}
As discussed in Section~\ref{sec:intro}, (\ref{eq:CH3}) has compactly supported solutions with zero contact slope. Here, we establish that energy minimizers share this property. We derive this result in one dimension, though one may extend the calculation to higher dimensions.

Consider an energy-minimizing solution $\bar{u}(x)$ with a support $x \in [\alpha,\beta]$. We compute the first variation with respect to changes in the endpoints in addition to the solution itself. Let
$\alpha = \bar{\alpha} + \delta{\alpha}$ , $\beta = \bar{\beta} + \delta{\beta}$, $u = \bar{u} + \delta{u}$. Computing the first variation of the energy, we find
\begin{equation}
\label{eq:dogface}
\delta E = \int_{\bar{\alpha}}^{\bar{\beta}} \frac{\delta E}{\delta u}\cdot \delta u\, dx + \frac{[u_x(\bar{\beta})]^2}{2}\cdot \delta \beta - \frac{[u_x(\bar{\alpha})]^2}{2}\cdot \delta \alpha,
\end{equation}
where
\begin{equation}
\frac{\delta E}{\delta u} = \frac{1}{2}u^2 - u - u_{xx} + W.
\end{equation}
In deriving (\ref{eq:dogface}), we have used the fact that $u$ vanishes at the endpoints $x=\bar{\alpha}$ and $x=\bar{\beta}$.
For $\delta E$ to vanish, it is necessary that $u_x(\bar{\alpha})=u_x(\bar{\beta})=0$. To see this, choose $\delta u = 0$ and suppose $u_x(\bar{\alpha})$ and $u_x(\bar{\beta})$ are nonzero. In this case, choosing $\delta \alpha > 0 $ would reduce the energy, as would choosing $\delta \beta < 0$. (The change in mass resulting from this shifting of the endpoints slightly into the support is a second-order effect.) However, this violates the assumption that $\bar{u}$ is a minimizer. Therefore, $u_x(\bar{\alpha})=u_x(\bar{\beta})=0$. Throughout much of the rest of our analysis of (\ref{eq:CH3}), we make use of the result we have just shown, namely that energy-minimizing solutions have zero contact slope.

From (\ref{eq:firstvariation}) and (\ref{eq:critpoint}), and for $W=0$, minimizers satisfy
\begin{equation}
\label{eq:steadyCH}
\frac{\delta E}{\delta u} = \frac{1}{2}u^2 - u - u_{xx} = \lambda.
\end{equation}
This differential equation has arisen previously in studies of the Cahn-Hilliard equation \cite{GriNov1995} and the thin film equation \cite{LauPug2000} and has solutions that can be expressed in terms of elliptic functions.
Multiply through by $u_x$, integrate once, and enforce that the energy minimizing solution has zero contact slope to obtain
\begin{equation}
\label{eq:chpp1}
\frac{u^3}{6} - \frac{u^2}{2} - \lambda u - \frac{u_x^2}{2} = 0.
\end{equation}
By symmetry, $u_x(0) = 0$ from which it follows that
\begin{equation}
u \left( \frac{u^2}{6} - \frac{u}{2} - \lambda \right) = 0,
\end{equation}
at the center of the clump solution. Therefore, we know the relationship between $\|u\|_\infty$ and $\lambda$, namely
\begin{equation}
\lambda = \frac{1}{6}\|u\|_\infty^2 - \frac{1}{2}\|u\|_\infty.
\end{equation}
We derive expressions for the size of the support of the solution $\|\supp(u)\|$ and for the mass $M$ in terms of $\|u\|_\infty$. For convenience, rearrange (\ref{eq:chpp1}) as
\begin{equation}
\left(\frac{du}{dx}\right)^2 = \frac{u^3}{3} - u^2 - 2\lambda u.
\end{equation}
For $\|\supp(u)\|$, we have
\begin{eqnarray}
\|\supp(u)\| & = & 2 \int_0^{\|\supp(u)\|/2} dx,\\
& = & 2 \int_0^{\|u\|_\infty} \frac{du}{du/dx},\\
& = & 2 \int_0^{\|u\|_\infty} \frac{du}{\sqrt{\frac{u^3}{3} - u^2 - 2\lambda u}},\\
& = & \frac{12}{\sqrt{9-3\|u\|_\infty}} \mathbb{K} \left( \sqrt{\frac{\|u\|_\infty}{3-\|u\|_\infty}} \right),
\end{eqnarray}
where $\mathbb{K}$ represents the complete elliptic integral of the first kind. For mass $M$ we have
\begin{eqnarray}
M & = & 2 \int_0^{\|\supp(u)\|/2} u\,dx,\\
& = & 2 \int_0^{\|u\|_\infty} \frac{u\,du}{du/dx},\\
& = & 2 \int_0^{\|u\|_\infty} \frac{u\,du}{\sqrt{\frac{u^3}{3} - u^2 - 2\lambda u}},\\
& = & 4 \sqrt{9-3\|u\|_\infty} \left[\mathbb{K} \left( \sqrt{\frac{\|u\|_\infty}{3-\|u\|_\infty}} \right) - \mathbb{E} \left( \sqrt{\frac{\|u\|_\infty}{3-\|u\|_\infty}} \right) \right],
\end{eqnarray}
where $\mathbb{E}$ represents the complete elliptic integral of the second kind. As $\|u\|_\infty$ increases from $0$ to $3/2$, the mass $M$ increases from $0$ to infinity, as seen in Figure~\ref{fig:varymass1d}(a).

Ideally, we would invert these relationships to solve for $\|u\|_\infty$ and $\|\supp(u)\|$ as a function of $M$, but this is cumbersome due to the elliptic integral functions. Therefore, as we did previously for the nonlocal model (\ref{eq:NL3}), we will analyze the small and large mass limits, which yield simple explicit expressions for $\|u\|_\infty$ and $\|\supp(u)\|$.

For small masses, neglect the nonlinear term in (\ref{eq:steadyCH}) to obtain
\begin{equation}
u + u_{xx} = -\lambda,
\end{equation}
As mentioned above, the solution that minimizes $E$ has zero contact angle and must also respect the mass constraint. Up to translation, the unique solution is
\begin{equation}
\label{eq:cosbump}
u(x) = \begin{cases} \frac{M}{2\pi}(1 + \cos x) & |x| \leq \pi \\ 0 & |x| \geq 0 \end{cases}.
\end{equation}
Thus, $\|u\|_\infty = M/\pi$ and $\|\supp(u)\| = 2\pi$.

For large masses, the $u_{xx}$ term in (\ref{eq:steadyCH}) is negligible and the solution is identical to the large mass limit of the nonlocal case considered above. That is, as $M \to \infty$, $\|u\|_\infty \to 3/2$ and $\|\supp(u)\| \to 2M/3$.

\subsection{Minimizers in two dimensions}

We briefly discuss the small and large mass limits in two dimensions. The small mass limit for the nonlocal model (\ref{eq:NL3}) is difficult to compute explicitly because of convolution with the kernel $Q_a \propto \exp{-|\vec{x}|}$. An alternative approach is to use a different kernel such as the Bessel kernel used in \cite{CarHuaMar2014,BurFetHua2014}. For our kernel, numerical explorations and a scaling argument both suggest that in the limit $M \to 0$, $\|\supp(u)\|$ and $\|u\|_\infty$ both approach zero in a fashion analogous to the problem in one dimension. The small mass limit for the Cahn-Hilliard model (\ref{eq:CH3}) is also analogous to its corresponding one dimensional case, satisfying
\begin{equation}
u + \nabla^2 u = -\lambda.
\end{equation}
The solution is
\begin{equation}
u(r) = \begin{cases} \displaystyle{\frac{M}{\pi r_*^2 } \left[1 - \frac{J_0(r)}{J_0(r_*)}\right]}  & r \leq r_* \\ 0 & r \geq r_* \end{cases},
\end{equation}
where the radius of the support, $r_*$, is the first zero of $J_1(r)$, and $J_{0}$ and $J_{1}$ are Bessel functions of the first kind. The support and peak density are
\begin{equation}
\|\supp(u)\| = \pi r_*^2, \quad \|u\|_\infty =  \displaystyle{\frac{M}{\pi r_*^2 } \left[1 - \frac{1}{J_0(r_*)}\right]}.
\end{equation}

The large mass calculations for two dimensions are nearly identical to those for one dimension. For (\ref{eq:NL3}) and (\ref{eq:CH3}) as $M \to \infty$, $\|u\|_\infty \to 3/2$ and the minimizer is a disc of density $u=3/2$, having radius
\begin{equation}
r = \sqrt{\frac{2M}{3\pi}}.
\end{equation}

We highlight two key results from this section. First, the Cahn-Hilliard model (\ref{eq:CH3}) approximates the nonlocal model (\ref{eq:NL3}) well, especially for large masses and away from an aggregation's boundaries. Second, in the limit of large mass, both models produce groups with an energetically preferred peak density of $3/2$. This value is the benchmark against which we address a biologically motivated question: \emph{can an external potential modeling the environment be used to reduce the peak density of an aggregation?}

\section{Energy minimizers in an external potential well}
\label{sec:gaussian}

With (\ref{eq:CH3}) established as a reasonable approximation of (\ref{eq:NL3}), at least for large enough masses and away from transition regions, we now use the degenerate Cahn-Hilliard model as a testbed to investigate the effects of the environment; that is, we now allow $W \neq 0$. More specifically, we focus on cases where $W$ is a potential well, and show that when this well has sufficiently tall and steep sides, it can be replaced by an approximate obstacle potential that is either zero or a sufficiently large, and in fact effectively infinite, constant. In this scenario, the support of a minimizer is constrained to avoid locations where $W$ is large. For the remainder of this paper, we simply refer to ``obstacle potentials,'' but remind the reader 
%to keep in mind 
that in our computational implementation, these are approximate obstacle potentials with large, rather than infinite, height.

The velocity in (\ref{eq:CH3}) only depends on the gradient of $W$. Furthermore, adding a constant to $W$ affects the energy (\ref{eq:CHenergy}) by adding a multiple of the conserved mass. Therefore, $W$ is arbitrary up to a constant. Without loss of generality, take the minimum value of $W$ to be zero. As a prototypical example, consider a Gaussian barrier placed at each end of the interval $[0,L]$, that is,
\begin{equation}
\label{eq:mothalovingaussian}
W(x) = \frac{A}{2 \sqrt{\pi} \sigma} \left( \exp{-x^2/4\sigma^2} + \exp{-(L-x)^2/4\sigma^2} \right),
\end{equation}
where $A$ measures the volume of the barriers and $\sigma$ is the characteristic width.

\begin{figure}[t]
\begin{tikzpicture}
\begin{axis}[xlabel=$x$,ylabel=$u$,y label style={rotate=-90},width=\textwidth,height=0.5\textwidth,xmin=0,xmax=10,ymin=0,ymax=3,enlargelimits=0]
\pgfplotstableread{data/gaussian1dvaryA.txt}\A;
\addplot[no marks,solid] table[x=x,y=A1] {\A};
\addplot[no marks,solid] table[x=x,y=A2] {\A};
\addplot[no marks,solid] table[x=x,y=A3] {\A};
\addplot[no marks,solid] table[x=x,y=A4] {\A};
\addplot[no marks,solid] table[x=x,y=A5] {\A};
\addplot[no marks,solid] table[x=x,y=A6] {\A};
\addplot[no marks,solid] table[x=x,y=A7] {\A};
\addplot[no marks,solid] table[x=x,y=A8] {\A};
\addplot[no marks,solid] table[x=x,y=A9] {\A};
\addplot[no marks,solid] table[x=x,y=A10] {\A};
\node at (rel axis cs:0.05,0.92) {$(a)$};
\end{axis}
\end{tikzpicture}
\begin{tikzpicture}
\begin{axis}[xlabel=$x$,ylabel=$u$,y label style={rotate=-90},width=\textwidth,height=0.5\textwidth,xmin=0,xmax=10,ymin=0,ymax=3,enlargelimits=0]
\pgfplotstableread{data/gaussian1dvarysigma.txt}\sigma;
\addplot[no marks,solid] table[x=x,y=sigma1] {\sigma};
\addplot[no marks,solid] table[x=x,y=sigma2] {\sigma};
\addplot[no marks,solid] table[x=x,y=sigma3] {\sigma};
\addplot[no marks,solid] table[x=x,y=sigma4] {\sigma};
\addplot[no marks,solid] table[x=x,y=sigma5] {\sigma};
\addplot[no marks,solid] table[x=x,y=sigma6] {\sigma};
\addplot[no marks,solid] table[x=x,y=sigma7] {\sigma};
\addplot[no marks,solid] table[x=x,y=sigma8] {\sigma};
\addplot[no marks,solid] table[x=x,y=sigma9] {\sigma};
\addplot[no marks,solid] table[x=x,y=sigma10] {\sigma};
\addplot[no marks,solid] table[x=x,y=sigma11] {\sigma};
\node at (rel axis cs:0.05,0.92) {$(b)$};
\end{axis}
\end{tikzpicture}
\caption{\label{fig:smoothpotential} Minimizers of the degenerate Cahn-Hilliard equation (\ref{eq:CH3}) with Gaussian potential barriers at the domain edges, as described by $W(x)$ in (\ref{eq:mothalovingaussian}). Domain length $L=10$ with $N=160$ computational grid points. Total mass is $M=15$. (a) Fixed Gaussian width $\sigma=L/40$ for potential volume $A$ varying between $1$ and $10$. As the amplitude increases, $u(0)$ and $u(L)$ are driven to zero and for sufficiently large $A$ there is a vacant segment near the potential barrier. (b) Fixed Gaussian amplitude $A=10$ for width $\sigma$ varying logarithmically between $0.0025$ and $0.25$. Here, $A$ is sufficiently large to drive $u$ to zero near the boundary. As $\sigma$ decreases, the vacant segment shrinks. The smallest few values of $\sigma$ to yield effectively nonzero $W$ only at a the edges of the computational domain; that is to say, $W$ is large at the domain endpoints and zero within numerical tolerance elsewhere. This choice pins the minimizer to be zero just at the endpoints, where a finite contact slope is observed.}
\end{figure}
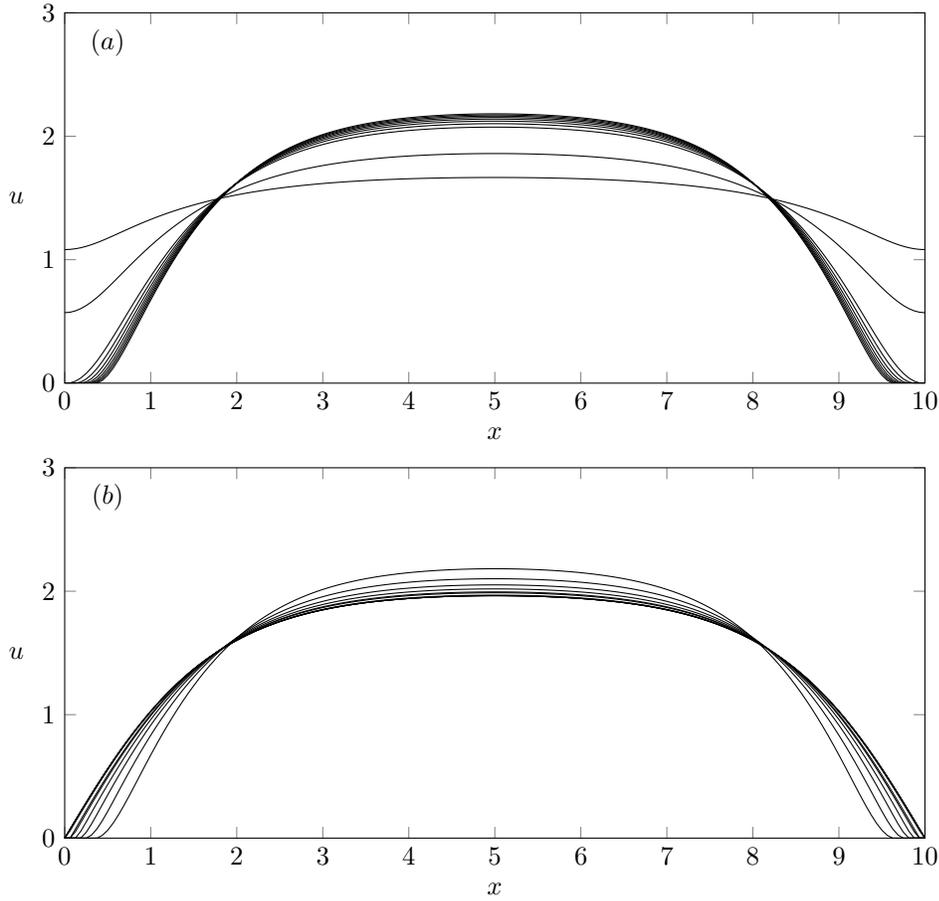

Figure~\ref{fig:smoothpotential} shows the ensuing minimizers of (\ref{eq:CH3}). Panel (a) varies the height $A$ with fixed width $\sigma = L/40$ on a domain of length $L=10$ with mass $M=15$. As $A$ increases, the density at the boundary decreases and is eventually driven to $u=0$. For sufficiently large $A$, $W$ drives the minimizer to zero over an interval, creating vacancies near the barriers. Panel (b) is similar, but varies $\sigma$ with fixed $A = 10$, which is sufficiently large to ensure $u=0$ at the boundary. As the barrier width $\sigma$ decreases, the vacant interval shrinks to a point, and $W$ pins the minimizer to $u=0$ at the boundaries. For the smallest few values of $\sigma$, $W$ is nonzero only at a the edges of the computational domain; that is to say, $W$ approaches infinity at the domain endpoints and is zero within numerical tolerance elsewhere.

To restate: as we decrease $\sigma$, the narrow Gaussian, which we have numerically under-resolved, approaches an obstacle potential, and the minimizer approaches one which is pinned to $u=0$ at the boundary. This limit suggests that we may replace our smooth potential $W$ with an approximate obstacle potential that is large on some subset of the gridpoints and zero elsewhere. In this limit, continuity of $W$ is lost. One consequence is that the analysis in Section~\ref{sec:ch1d} demonstrating zero contact slope for minimizers no longer holds. Indeed, it is apparent in Figure~\ref{fig:smoothpotential}(b) that minimizers may have nonzero slope where they are pinned by the potential at the domain boundary.

%%%%%%%%%%%%%%%%%%%%%%%%%%%%%%%%%%%%%%
%%%%%%%%%%%%%%%%%%%%%%%%%%%%%%%%%%%%%%

\begin{figure}[t]
\begin{center}
\begin{tikzpicture}
\begin{axis}[xlabel=$x$,ylabel=$y$,y label style={rotate=-90},width = 0.48 \textwidth,height = 0.48\textwidth,xmin=0,xmax=3,ymin=0,ymax=3]
\addplot graphics [xmin=0,xmax=3,ymin=0,ymax=3]{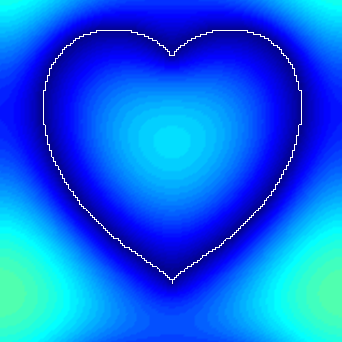};
\end{axis}
\end{tikzpicture}
\begin{tikzpicture}
\begin{axis}[xlabel=$x$,ylabel=$y$,y label style={rotate=-90},width = 0.48 \textwidth,height = 0.48\textwidth,xmin=0,xmax=3,ymin=0,ymax=3, colorbar, colormap/jet, point meta min = 0, point meta max =5]
\addplot graphics [xmin=0,xmax=3,ymin=0,ymax=3]{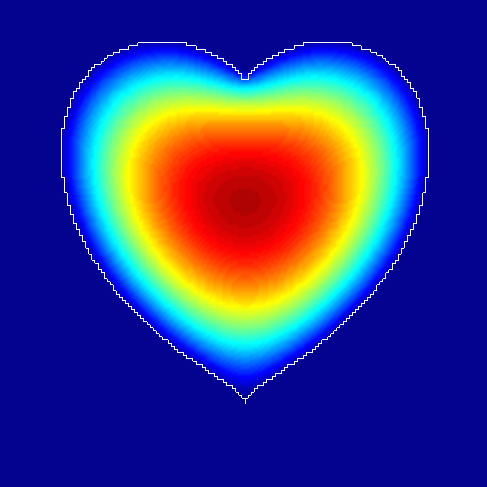};
\end{axis}
\end{tikzpicture}
\end{center}
\caption{\label{fig:heart2d} Minimizers of the degenerate Cahn-Hilliard equation (\ref{eq:CH3}) for two obstacle potentials $W(\vec{x})$. The domain has sides of length $L=3$ with $N=160$ computational grid points along each axis. Total mass is $M=9$. (a) $W(\vec{x}) = 10^3$ on the heart-shaped curve, and $W=0$ elsewhere. The potential pins the minimizers to $u=0$ along this curve. (b) Similar to (a), but the potential is now nonzero on and exterior to the curve, creating a heart-shaped well. In the absence of a potential, the minimizer is a constant, $u(\vec{x})=1$. In (a), $W$ drives the peak density higher, to $\|u\|_\infty \approx 2.2$. In (b), $W$ drives the peak density even higher, to $\|u\|_\infty \approx 4.7$.}
\end{figure}

The strategy of adopting an obstacle potential is effective in two dimensions as well. Figure~\ref{fig:heart2d} shows a minimizer of (\ref{eq:CH3}) with mass $M=9$ on a domain with sides of length $L=3$. In the absence of an external potential, the minimizer has constant density $u(\vec{x}) = 1$. Panel (a) takes the potential $W(\vec{x}) = 10^3$ at computational gridpoints that trace out the white heart-shaped curve, and $W=0$ elsewhere. The potential drives the minimizer to $u=0$ along this curve, apparent as the dark blue region, and the peak density is $\|u\|_\infty \approx 2.2$. Panel (b) is similar, but takes $W(\vec{x}) = 10^3$ on and exterior to the curve, creating a heart-shaped well. The potential confines the minimizer to the well, drives it to zero along the boundary, and pushes the peak density even higher, to $\|u\|_\infty \approx 4.7$.

%%%%%%%%%%%%%%%%%%%%%%%%%%%%%%%%%%%%%%
%%%%%%%%%%%%%%%%%%%%%%%%%%%%%%%%%%%%%%

The next section focuses on cases where $W$ is an obstacle potential, and examines strategies for reducing the peak density by engineering $W$.

\section{The effect on peak density of random versus periodic potentials}
\label{sec:randper}

Our biological motivation for this study is the dynamics of locust populations. Intuitively, one might think that the distribution of many organisms in the wild is affected by the distribution of resources on which they subsist, and indeed, this issue is part of the rich field of biogeography. The way that resources shape the distribution of locusts is intimately tied to locusts' phase polyphenism. Phase polyphenism refers to individuals of the same genotype manifesting different phenotypes \cite{AppHei1999,PenSim2009}. For locusts, a pivotal phenotypic difference is the insect's social behavior, or lack thereof. A locust may exist in a solitarious phase in which it seeks isolation or a gregarious phase in which it seeks other locusts. Locusts in the gregarious phase may form large, migratory swarms that decimate crops. Therefore, dense social aggregations on the ground are of interest as they are precursors to dangerous flying groups.

Previous work \cite{TopDOrEde2012} identifies a population density threshold above which a collective transition to a dangerous, gregarious group takes place. Thus, one way to prevent airborne swarms might be to suppress the density of locusts on the ground, so that the group never reaches the critical threshold. Because social locusts display taxis not only to each other, but to food, one can ask whether there exist external potentials $W$, modeling food sources, which serve to minimize the peak density $\|u\|_\infty$ of the population. We investigate this problem within the framework of (\ref{eq:CH3}). The locust model in \cite{TopDOrEde2012} includes additional effects -- crucially, the phase change from solitarious to gregarious locusts -- but an understanding of the interaction between locusts and food sources even in the absence of phase change is an appropriate starting point for investigation.

\begin{figure}[t]
\begin{tikzpicture}
\begin{groupplot}[group style={group size=3 by 3, horizontal sep = 45pt, vertical sep = 40 pt},xlabel=$x$, ylabel=$u$, xmin = 0, xmax = 37.7, ymin = 0, ymax = 2.5, width = 0.34\textwidth, xtick={0,18.85,37.7}, xticklabels={$0$,$6\pi$,$12\pi$}, xtick align = outside, xtick pos = left]

\nextgroupplot
\pgfplotstableread{data/subdivisions.txt}\data;
\addplot[mark=none, solid] table[x=x, y=M1n1] {\data};
\node at (rel axis cs:0.09,0.88) {$(a)$};
\nextgroupplot
\pgfplotstableread{data/subdivisions.txt}\data;
\addplot[mark=none, solid] table[x=x, y=M1n2] {\data};
\node at (rel axis cs:0.09,0.88) {$(b)$};
\nextgroupplot
\pgfplotstableread{data/subdivisions.txt}\data;
\addplot[mark=none, solid] table[x=x, y=M1n3] {\data};
\node at (rel axis cs:0.09,0.88) {$(c)$};

\nextgroupplot
\pgfplotstableread{data/subdivisions.txt}\data;
\addplot[mark=none, solid] table[x=x, y=M2n1] {\data};
\node at (rel axis cs:0.09,0.88) {$(d)$};
\nextgroupplot
\pgfplotstableread{data/subdivisions.txt}\data;
\addplot[mark=none, solid] table[x=x, y=M2n2] {\data};
\node at (rel axis cs:0.09,0.88) {$(e)$};
\nextgroupplot
\pgfplotstableread{data/subdivisions.txt}\data;
\addplot[mark=none, solid] table[x=x, y=M2n3] {\data};
\node at (rel axis cs:0.09,0.88) {$(f)$};

\nextgroupplot
\pgfplotstableread{data/subdivisions.txt}\data;
\addplot[mark=none, solid] table[x=x, y=M3n1] {\data};
\node at (rel axis cs:0.09,0.88) {$(g)$};
\nextgroupplot
\pgfplotstableread{data/subdivisions.txt}\data;
\addplot[mark=none, solid] table[x=x, y=M3n2] {\data};
\node at (rel axis cs:0.09,0.88) {$(h)$};
\nextgroupplot
\pgfplotstableread{data/subdivisions.txt}\data;
\addplot[mark=none, solid] table[x=x, y=M3n3] {\data};
\node at (rel axis cs:0.09,0.88) {$(i)$};

\end{groupplot}
\end{tikzpicture}
\caption{\label{fig:subdivisions} Minimizers of the Cahn-Hilliard model (\ref{eq:CH3}) on a domain of length $L=12\pi$ for varying mass $M$ and varying numbers of square wells in the external potential $W(x)$. (a-c) $M=\pi$. (d-f) $M=4\pi$. (g-i) $M=16\pi$.  In the first column, there is one square well, an arrangement we use in analogy to an isolated plot of planted land. In the second column, there are eight square wells, and in the third column there are 24. At low and intermediate $M$ (top and middle rows), applying a large number of square wells in $W(x)$ to the fixed domain reduces the peak density $\|u\|_\infty$, as seen in (c) and (f). In contrast, at large $M$ (bottom row), increasing the number of square wells applied via $W(x)$ increases $\|u\|_\infty$. For the numerical computations we use $N=744$ grid points.}
\end{figure}
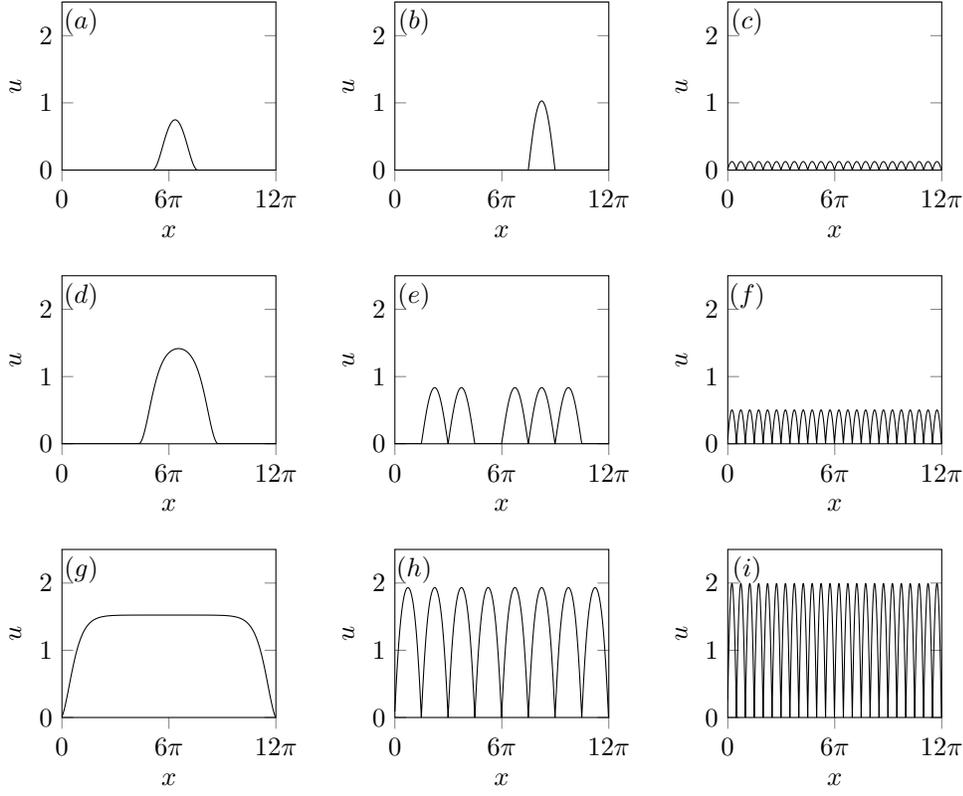

Figure~\ref{fig:subdivisions} reveals some of the intricacies of the problem. Consider a one dimensional domain and obstacle potentials $W(x)$ describing a sequence of square wells with dividers between them, intended to model cropland divided into plots. Each panel shows a minimizer of (\ref{eq:CH3}). At low and moderate masses $M$ (top and middle rows), applying a large number of square wells in $W(x)$ to the fixed domain reduces the peak density $\|u\|_\infty$, whereas at high masses (bottom row) it has the opposite effect. Furthermore, at low and moderate densities, social aggregation can cause the mass to clump in a subset of the wells, as seen in panels (b) and (e), which may serve to either reduce or increase peak density.

%%%%%%%%%%%%%%%%%%%%%%%%%%%%%%%%%%%%%%
%%%%%%%%%%%%%%%%%%%%%%%%%%%%%%%%%%%%%%

In summary, the goal of this section is to choose external potentials $W$ to minimize the peak density $\|u\|_\infty$ using the Cahn-Hilliard model (\ref{eq:CH3}) as a testbed. For tractability, we restrict attention to the class of obstacle potentials. The next three subsections focus on a one dimensional domain with periodic $W$. In the the final subsection, we compare results for periodic potentials with those for random potentials in both one and two dimensions. Random potentials tend to increase peak density $\|u\|_\infty$ while periodic potentials can decrease it.

\subsection{Energy minimizers in a single square well}

For the Cahn-Hilliard model (\ref{eq:CH3}), when $\|\supp(u)\|$ for a given mass on an infinite domain is less than the well width $\ell$, the minimizer is identical to that found in Section~\ref{sec:minimizers}. However, when the support is larger, the minimizer feels the boundaries of the square well. For sufficiently large $W$ at the edge of the well (that is, sufficient contrast between desert and planted regions), $u$ will be driven to zero.

From (\ref{eq:firstvariation}) and (\ref{eq:critpoint}), minimizers satisfy
\begin{equation}
\label{eq:onewell}
\frac{\delta E}{\delta u} = \frac{1}{2}u^2 - u - u_{xx} = \lambda,
\end{equation}
subject to the mass constraint and the added restriction that $u=0$ at the edges of the well because of the external potential. An  implicit solution for $u$ as a function of $\lambda$, mass $M$, and well size $\ell$ could be found in terms of elliptic integrals.  However it is simpler to consider small and large mass limits, as in Section~\ref{sec:minimizers}, and augment our calculations with numerical computations. First, we calculate the energy $E$ of the minimizing solution for several cases depending on $M$ and $\ell$.

Case IA: Small $M$, $\ell < 2\pi$. In Section~\ref{sec:minimizers}, we showed that for an infinite domain, the small $M$ solutions have support of size $2\pi$. We expect that for our present case, the solution will occupy the entire domain. To examine the small mass limit, linearize (\ref{eq:onewell}) to obtain the boundary value problem
\begin{equation}
- u - u_{xx} = \lambda, \quad u(0)=u(\ell)=0,
\end{equation}
in addition to the mass constraint. The solution is
\begin{equation}
u(x) = M \frac{\cos(x-\ell/2) - \cos(\ell/2)}{2\sin(\ell/2) - \ell \cos(\ell/2)}.
\end{equation}
Direct computation via substitution into (\ref{eq:CHenergy}) yields the energy for this solution,\begin{equation}
\label{eq:Ewellsmallmass1}
E = \frac{M^2}{2} \frac{\cos(\ell/2)}{2\sin(\ell/2)-\ell\cos(\ell/2)} + \order(M^4).
\end{equation}
Crucially, the energy changes from positive to negative as $\ell$ increases through $\pi$.

Case IB: Small $M$, $\ell > 2\pi$. The results are identical to the infinite interval result (so long as the support of the solution is less than $\ell$). Substituting (\ref{eq:cosbump}) into (\ref{eq:CHenergy}), we find the energy
\begin{equation}
E = - \frac{M^2}{4\pi}+ \order(M^4).
\end{equation}

Case II: Large $M$. Numerical simulation shows the density profile to be constant in the potential well except for a small boundary layer near the sides of the well of thickness $\order(\sqrt{\ell/M})$. Exploiting this fact, we have
\begin{equation}
u(x) = M/\ell,
\end{equation}
and thus the energy is
\begin{equation}
\label{eq:Ewelllargemass}
E = \frac{M^3}{6 \ell^2} + \order\left(\frac{M^{5/2}}{\ell^{5/2}}\right).
\end{equation}

\subsection{Energy per unit mass in a single well}

Momentarily, we will make predictions about energy minimizers in multiple wells. To minimize the total energy we will show that a system with multiple wells distributes mass among those wells to minimize the energy per unit mass. Thus, as another preliminary step, we use the results of the previous subsection to calculate the energy per unit mass in a single well. Define the energy per unit mass,
\begin{equation}
\mathcal{E}(M) = E(M)/M.
\end{equation}
From the previous subsection, we can determine the behavior of $\mathcal{E}(M)$ for the limiting cases of small and large $M$. We will see that $\mathcal{E}(M)$ may have a minimum, which will be a crucial characteristic when we consider the case of multiple wells. We refer to the value of $M$ that minimizes $\mathcal{E}(M)$ as $M^*$.

%%%%%%%%%%%%%%%%%%%%%%%%%%%%%%%%%%%%%%
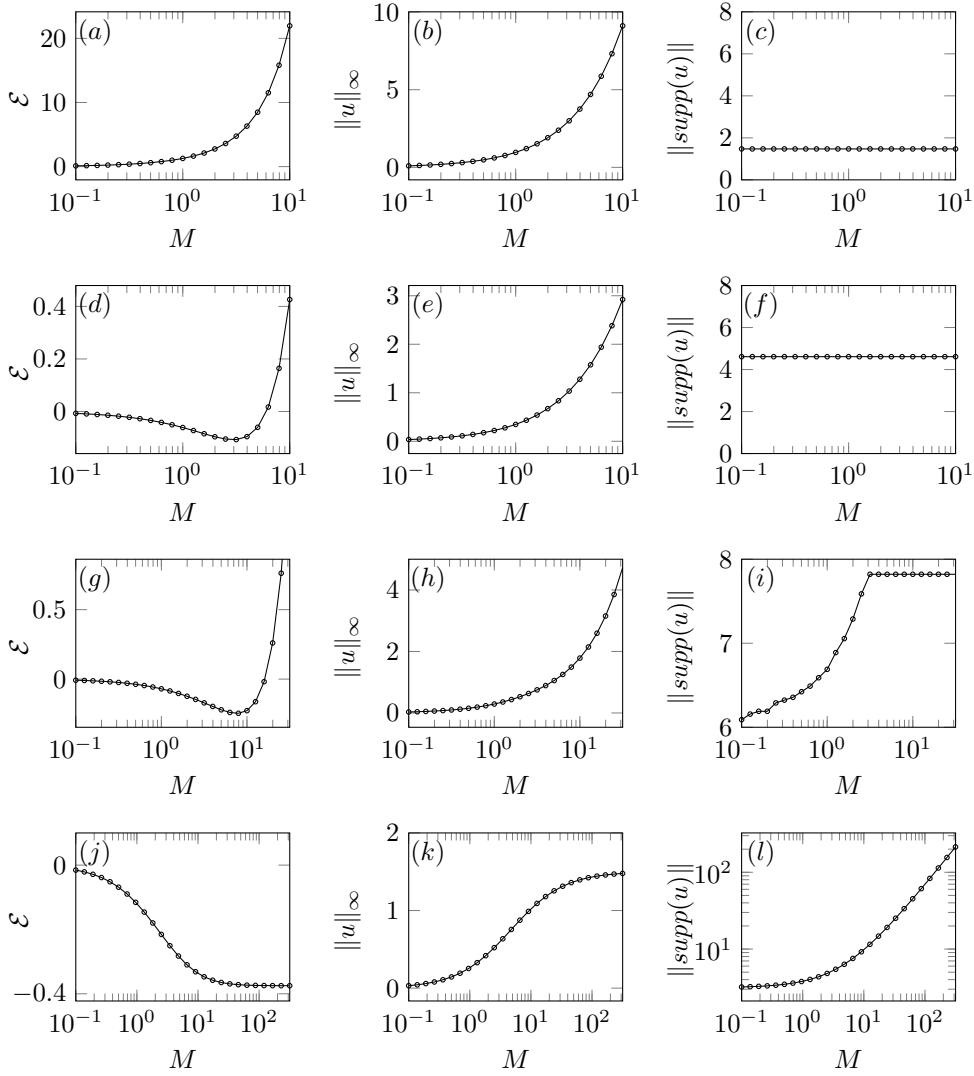
\begin{figure}[t]
\begin{tikzpicture}
\pgfplotsset{y label style={at={(axis description cs:-0.18,.48)},anchor=south}}
\begin{groupplot}[group style={group size=3 by 4, horizontal sep = 45pt, vertical sep = 40 pt},xlabel=$M$,width = 0.34\textwidth, xmode=log]
\nextgroupplot[ylabel=$\mathcal{E}$, xmin = 0.1, xmax = 10]
\pgfplotstableread{data/Lpiover2.txt}\data;
\addplot[mark=o, mark size = 0.85, solid] table[x= M, y=EperM] {\data};
\node at (rel axis cs:0.09,0.88) {$(a)$};
\nextgroupplot[ylabel=$\|u\|_\infty$, xmin = 0.1, xmax = 10]
\pgfplotstableread{data/Lpiover2.txt}\data;
\addplot[mark=o, mark size = 0.85, solid] table[x= M, y=umax] {\data};
\node at (rel axis cs:0.09,0.88) {$(b)$};
\nextgroupplot[ylabel=$\|supp(u)\|$, xmin = 0.1, xmax = 10, ymin = 0, ymax = 8]
\pgfplotstableread{data/Lpiover2.txt}\data;
\addplot[mark=o, mark size = 0.85, solid] table[x= M, y=usupp] {\data};
\node at (rel axis cs:0.09,0.88) {$(c)$};

\nextgroupplot[ylabel=$\mathcal{E}$, xmin = 0.1, xmax = 10]
\pgfplotstableread{data/L3piover2.txt}\data;
\addplot[mark=o, mark size = 0.85, solid] table[x= M, y=EperM] {\data};
\node at (rel axis cs:0.09,0.88) {$(d)$};
\nextgroupplot[ylabel=$\|u\|_\infty$, xmin = 0.1, xmax = 10]
\pgfplotstableread{data/L3piover2.txt}\data;
\addplot[mark=o, mark size = 0.85, solid] table[x= M, y=umax] {\data};
\node at (rel axis cs:0.09,0.88) {$(e)$};
\nextgroupplot[ylabel=$\|supp(u)\|$, xmin = 0.1, xmax = 10, ymin = 0, ymax = 8]
\pgfplotstableread{data/L3piover2.txt}\data;
\addplot[mark=o, mark size = 0.85, solid] table[x= M, y=usupp] {\data};
\node at (rel axis cs:0.09,0.88) {$(f)$};

\nextgroupplot[ylabel=$\mathcal{E}$, xmin = 0.1, xmax = 31.6]
\pgfplotstableread{data/L5piover2.txt}\data;
\addplot[mark=o, mark size = 0.85, solid] table[x= M, y=EperM] {\data};
\node at (rel axis cs:0.09,0.88) {$(g)$};
\nextgroupplot[ylabel=$\|u\|_\infty$, xmin = 0.1, xmax = 31.6,  ymax = 5]
\pgfplotstableread{data/L5piover2.txt}\data;
\addplot[mark=o, mark size = 0.85, solid] table[x= M, y=umax] {\data};
\node at (rel axis cs:0.09,0.88) {$(h)$};
\nextgroupplot[ylabel=$\|supp(u)\|$, xmin = 0.1, xmax = 31.6, ymin = 6, ymax = 8, ytick={6,7,8}]
\pgfplotstableread{data/L5piover2.txt}\data;
\addplot[mark=o, mark size = 0.85, solid] table[x= M, y=usupp] {\data};
\node at (rel axis cs:0.09,0.88) {$(i)$};

\nextgroupplot[ylabel=$\mathcal{E}$, xmin = 0.1, xmax = 317, ymax = 0.1, ytick={-0.4,0}]
\pgfplotstableread{data/Linf.txt}\data;
\addplot[mark=o, mark size = 0.85, solid] table[x= M, y=EperM] {\data};
\node at (rel axis cs:0.09,0.88) {$(j)$};
\nextgroupplot[ylabel=$\|u\|_\infty$, xmin = 0.1, xmax = 317, ymax = 2, ytick = {0,1,2}]
\pgfplotstableread{data/Linf.txt}\data;
\addplot[mark=o, mark size = 0.85, solid] table[x= M, y=umax] {\data};
\node at (rel axis cs:0.09,0.88) {$(k)$};
\nextgroupplot[ylabel=$\|supp(u)\|$, xmin = 0.1, xmax = 317, ymode = log]
\pgfplotstableread{data/Linf.txt}\data;
\addplot[mark=o, mark size = 0.85, solid] table[x= M, y=usupp] {\data};
\node at (rel axis cs:0.09,0.88) {$(l)$};

\end{groupplot}
\end{tikzpicture}
\caption{\label{fig:energy1d} Properties of minimizers of the Cahn-Hilliard model (\ref{eq:CH3}) with a single square well of width $\ell$ used for the external potential $W(x)$. Columns are the energy per unit mass $\mathcal{E}(M)$, the peak density $\|u\|_\infty$, and the size of the support $\|\supp(u)\|$, all as a function of mass $M$. (a-c) Case I,  $\ell = \pi/2$. Here, $\mathcal{E}(M)$ and $\|u\|_\infty$ increases monotonically and $\supp(u)$ fills the entire domain. (d-f) Case II,  $\ell = 3 \pi/2$. Here, $\mathcal{E}(M)$ reaches a minimum at $M=M^*$, $\|u\|_\infty$ increases monotonically and $\supp(u)$ fills the entire domain. (g-i) Case III, $\ell=5 \pi/2$. Here, $\mathcal{E}(M)$ reaches a minimum at $M=M^*$, $\|u\|_\infty$ increase monotonically and $\supp(u)$ widens and eventually fills the entire domain. (j-l) Case IV, $\ell$ chosen to be effectively infinite, much larger than $\supp(u)$. Here, $\mathcal{E}(M)$ decreases monotonically to $-3/8$ and $\|u\|_\infty$ increases monotonically to $3/2$. Additionally, $\supp(u)$ widens without bound and, asymptotically, grows linearly.}
\end{figure}
%%%%%%%%%%%%%%%%%%%%%%%%%%%%%%%%%%%%%%

Case I: $\ell < \pi$. A sample computation of $\mathcal{E}(M)$ is shown in Figure~\ref{fig:energy1d}(a). It is monotonically increasing. This result is consistent with the small mass result (\ref{eq:Ewellsmallmass1}), from which it follows that
\begin{equation}
\label{eq:gah1}
\mathcal{E}(M) = \frac{M}{2} \frac{\cos(\ell/2)}{2\sin(\ell/2)-\ell\cos(\ell/2)},
\end{equation}
which is linearly increasing with $M$. Also, from the large mass result (\ref{eq:Ewelllargemass}), 
\begin{equation}
\label{eq:gah2}
\mathcal{E}(M) = \frac{M^2}{6 \ell^2} ,
\end{equation} 
which is again increasing.  We conclude that $M^* = 0$, that is, having the minimum possible mass in the well minimizes $\mathcal{E}(M)$.

Case II: $\pi < \ell < 2\pi$. A sample calculation of $\mathcal{E}(M)$ is shown in Figure~\ref{fig:energy1d}(d). Relations (\ref{eq:gah1}) and (\ref{eq:gah2}) still hold, but crucially, the coefficient in the small mass limit is negative rather than positive, as it was in Case I. Thus, we expect $M^*$ to occur between the decreasing behavior at small mass and the increasing behavior at large mass, as evidenced in the figure.

Case III: $\ell > 2\pi$ (but finite). Recall that for an infinite domain and in the limit of small mass, the minimizer is given by (\ref{eq:cosbump}) and the support is $2\pi$. It is not surprising, therefore, that the same small mass result holds for intervals wider than $2\pi$. Plugging the minimizer into the energy and dividing by $M$ yields
\begin{equation}
\label{eq:gah3}
\mathcal{E}(M) = -\frac{M}{4 \pi},
\end{equation}
which is decreasing. As the mass increases, the support widens as well, and eventually fills the domain, as shown in Figure~\ref{fig:energy1d}(i). For sufficiently large mass, the density is nearly constant with $u(x) \approx M/\ell$ (except at the boundaries) and the approximation (\ref{eq:gah2}) still holds. As in Case II above, $\mathcal{E}(M)$ is decreasing at small mass, increasing at large mass and reaches a minimum at $M^*$ between these two regimes.

Case IV: $\ell$ large (effectively infinite). The small mass approximation (\ref{eq:gah3}) still holds. As $M$ increases, the minimizer follows the branch of solutions found for the infinite interval. Moreover, by the energy argument in Section \ref{sec:minimizers}, the solution approaches a single rectangle of height $u=3/2$, having, by direct calculation,
\begin{equation}
\label{eq:gah4}
\mathcal{E}(M) = -\frac{3}{8}.
\end{equation}
Thus, $\mathcal{E}(M)$ decreases monotonically but approaches this constant, and so $M^*$ is effectively infinite. It will be energetically favorable for the mass to form a clump of density $u=3/2$ inside a single well as was seen previously.

\subsection{Energy minimizers in multiple wells}

We now take the external potential $W$ to consist of a periodic sequence of square wells of width $\ell$. The key question is whether the mass will distribute itself equally amongst the wells or whether it is energetically preferred to concentrate in a subset of the wells. We use the results from the previous subsections to answer this question.

More concretely, suppose that there are $n$ equal wells in the domain and that for each well, $E(m)$ is the energy associated with a minimizer of mass $m$ located in that well. The total energy for a distribution of masses $m_i>0$ is
\begin{equation}
E(M) = \sum_{i=1}^n E(m_i),
\end{equation}
where
\begin{equation}
\sum_{i=1}^n m_i = M,
\end{equation}
because total mass is fixed. To proceed, consider an \emph{ansatz} where the mass is equally distributed between $k$ of the $n$ wells. In this case,
\begin{equation}
E(M) = k E(M/k) = \frac{M}{\mu} E(\mu) = M \mathcal{E}(\mu),
\end{equation}
where $\mu$ is the mass contained in each occupied well. From this equation, we deduce that a necessary condition to minimize $E(M)$ is to minimize the energy per unit mass $\mathcal{E}(\mu)$ over admitted values of $\mu$.

For the sake of argument, suppose one can choose $\mu = M^*$. In this case, $E(M)$ is globally minimized. In practice, the values of $\mu$ are quantized since $\mu = M/k$ where $k$ is a positive integer less than or equal to $n$. For large enough $n$, the system appears to choose $\mu$ that approximates $M^*$ well.

Now apply these ideas to the numerical results in Figure~\ref{fig:subdivisions}. For panels (a,d,g) in the first column, there is one square well, that is, $\ell = L = 12\pi$. For panels (b,e,h) in the second column, there are eight square wells, that is $\ell = L/8 = 3 \pi/2$, for which we have estimated $M^* \approx 2.97$ from the data in Figure~\ref{fig:energy1d}(d). In panel (b), as $M \approx M^*$, the mass fills a single well. In panel (e), as $4 M^* < M < 5M^*$, the mass fills five of the wells. In panel (h), as $M > 8 M^*$, the mass equipartitions among all eight wells. Finally, for panels (c,f,i) in the third column, there are 24 square wells, that is, $\ell = L/24 = \pi/2$ for which $M^* = 0$. Therefore, for all values of $M$, mass equipartitions among all wells.

\subsection{Comparing random and periodic potentials}

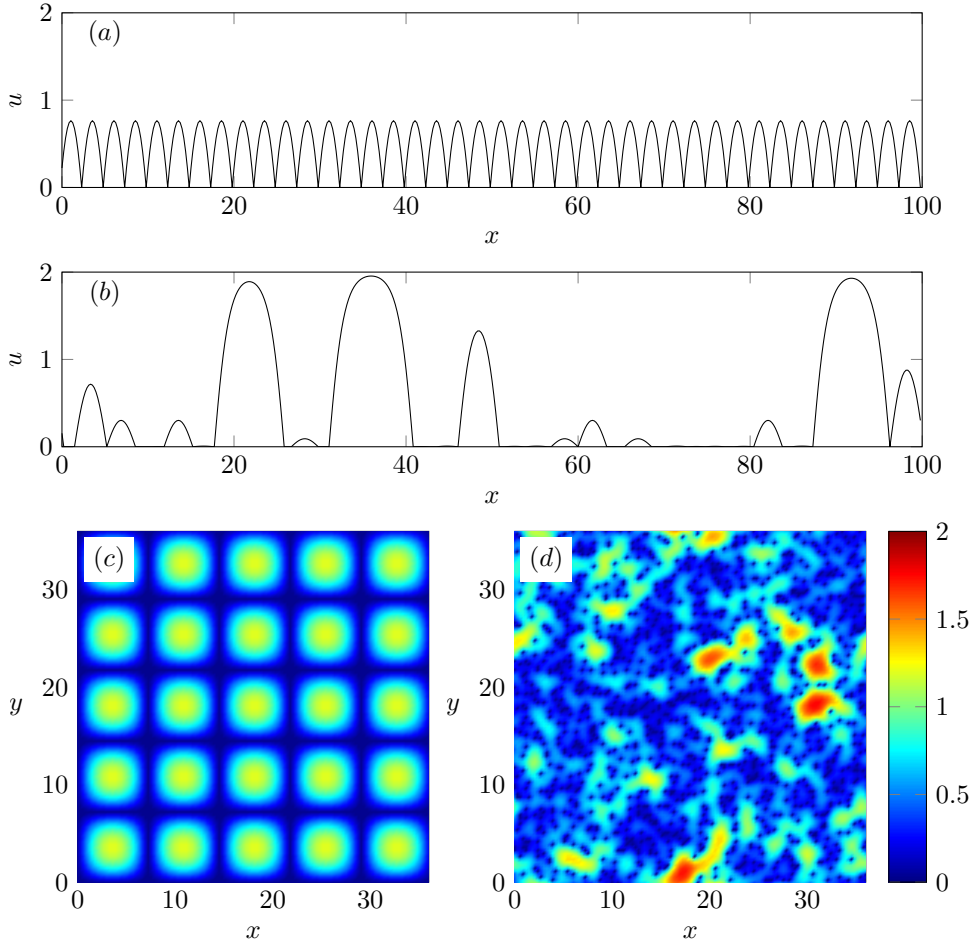
\begin{figure}[t]
\begin{tikzpicture}
\begin{axis}[xlabel=$x$,ylabel=$u$,y label style={rotate=0},width=\textwidth,height=0.3\textwidth,xmin=0,xmax=100,ymin=0,ymax=2,enlargelimits=0,xtick={0,20,40,60,80,100},ytick={0,1,2}]
\pgfplotstableread{data/periodic1dsnapshot.txt}\periodiconed;
\addplot[no marks] table[x=x,y=u] {\periodiconed};
\node at (rel axis cs:0.05,0.88) {$(a)$};
\end{axis}
\end{tikzpicture}
\begin{tikzpicture}
\begin{axis}[xlabel=$x$,ylabel=$u$,y label style={rotate=0},width=\textwidth,height=0.3\textwidth,xmin=0,xmax=100,ymin=0,ymax=2,enlargelimits=0,xtick={0,20,40,60,80,100},ytick={0,1,2}]
\pgfplotstableread{data/random1dsnapshot.txt}\randomoned;
\addplot[no marks] table[x=x,y=u] {\randomoned};
\node at (rel axis cs:0.05,0.88) {$(b)$};
\end{axis}
\end{tikzpicture}
\begin{tikzpicture}
\begin{axis}[xlabel=$x$,ylabel=$y$,y label style={rotate=-90},width = 0.48 \textwidth,height = 0.48\textwidth,xmin=0,xmax=36,ymin=0,ymax=36]
\addplot graphics [xmin=0,xmax=36,ymin=0,ymax=36]{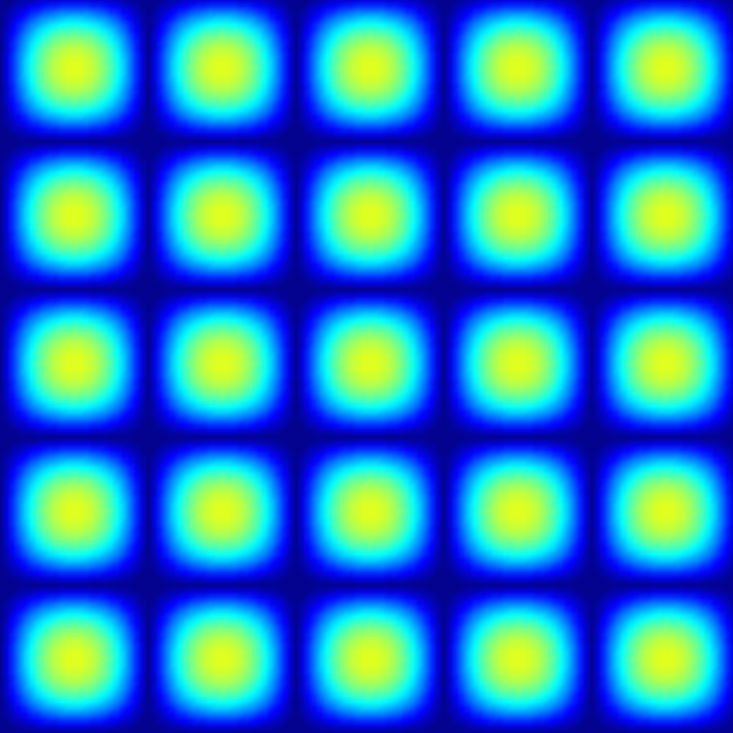};
\node at (rel axis cs:0.09,0.92) [fill = white] {$(c)$};
\end{axis}
\end{tikzpicture}
\begin{tikzpicture}
\begin{axis}[xlabel=$x$,ylabel=$y$,y label style={rotate=-90},width = 0.48 \textwidth,height = 0.48\textwidth,xmin=0,xmax=36,ymin=0,ymax=36, colorbar, colormap/jet, point meta min = 0, point meta max =2]
\addplot graphics [xmin=0,xmax=36,ymin=0,ymax=36]{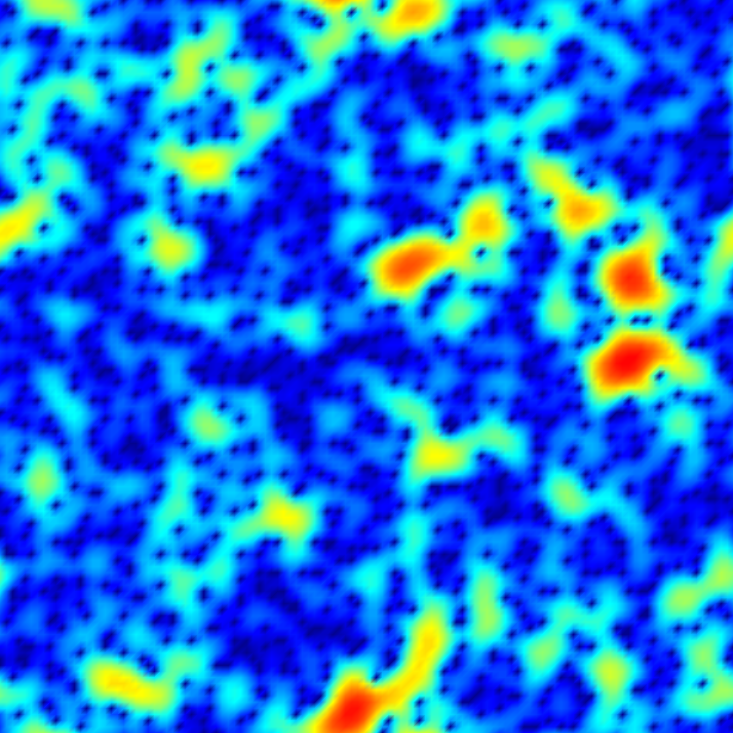};
\node at (rel axis cs:0.09,0.92) [fill = white] {$(d)$};
\end{axis}
\end{tikzpicture}
\caption{\label{fig:randpersnapshots} Example minimizers of (\ref{eq:CH3}) for random and periodic obstacle potentials $W(\vec{x})$. For these potentials, we fix $\phi$, the volume fraction of the computational domain in which the environmental potential $W$ is effectively infinite (here, $10^4$). Elsewhere, $W=0$. These choices of W model, respectively, regions barren and lush with vegetation. (a,b) Minimizers in one dimension for 
$M=50$ on a domain of length $L=100$ with $480$ computational gridpoints and $\phi = 1/12 \approx 0.083$. For the periodic potential used in (a), mass equipartitions into all 40 potential wells and $\|u\|_\infty \approx 0.76$. For the randomly distributed potential used in (b), some wells are left vacant and the peak density increases to $\|u\|_\infty \approx 1.95$. (c,d) Minimizers in two dimensions for $M=648$ on a domain with sides of length $L=36$ with $N=120$ computational grid points along each axis and $\phi = 47/576 \approx 0.082$. For the periodic potential used in (c), mass equipartitions into all 25 potential wells in the five-by-five grid, and $\|u\|_\infty \approx 1.2$. For the randomly distributed potential used in (d), there exist hot spots with peak densities as high as $\|u\|_\infty \approx 1.7$.}
\end{figure}

We now demonstrate that a periodic pattern of resources in the environment can reduce peak density below what might be seen with a random distribution of resources. This effect is evident in Figure~\ref{fig:randpersnapshots}, which compares these two cases in one and two dimensions. For all four panels, the average density is $u = 1/2$, which is below the peak density of $3/2$ seen in minimizers in the absence of an external potential (see Section \ref{sec:minimizers}). Additionally, we fix $\phi$, the volume fraction of the computational domain on which the obstacle potential $W$ is effectively infinite (here, $10^4$) and we take $W=0$ elsewhere. These choices of W model, respectively, regions barren and lush with vegetation. In Figure~\ref{fig:randpersnapshots}, we set $\phi \approx 0.08$.

First, consider the one dimensional case. In panel (a), mass equipartitions among all 40 available potential wells, and $\|u\|_\infty \approx 0.76$. For the randomly distributed potential used in (b), some wells are left vacant and the peak density increases to $\|u\|_\infty \approx 1.95$. Now, for two dimensions, panel (c) demonstrates that mass equipartitions among all 25 potential wells in the five-by-five grid, and $\|u\|_\infty \approx 1.2$. However, for the randomly distributed potential used in (d), there exist hot spots with peak densities as high as $\|u\|_\infty \approx 1.7$.

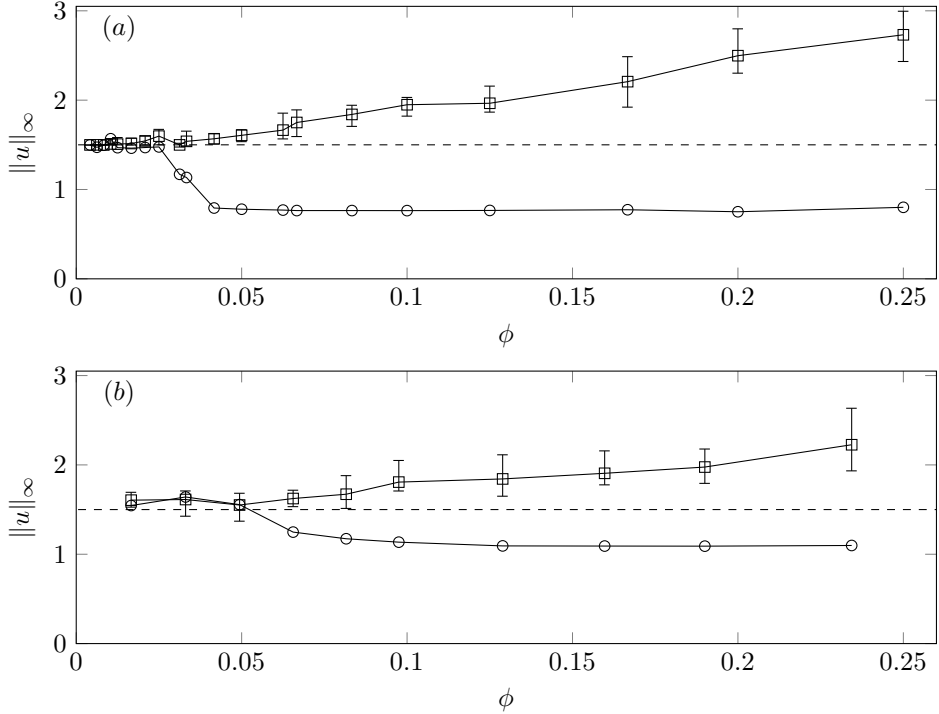
\begin{figure}[t]
\begin{tikzpicture}
\begin{axis}[xlabel=$\phi$,ylabel=$\|u\|_\infty$,y label style={rotate=0},width=\textwidth,height=0.4\textwidth,xmin=0,xmax=0.26,ymin=0,ymax=3.05,enlargelimits=0, xticklabel style={/pgf/number format/fixed}, xtick = {0,0.05,0.1,0.15,0.2,0.25}]
\pgfplotstableread{data/periodic1d.txt}\periodiconed;
\addplot[mark=o] table[x=phi,y=umax] {\periodiconed};
\pgfplotstableread{data/randomsummary1d.txt}\randomoned;
\addplot[mark = square, error bars/.cd, y dir = both, y explicit] table[x = phi, y = umaxmean, y error plus = eplus, y error minus = eminus]{\randomoned};
\addplot[dashed, no marks]{3/2};
\node at (rel axis cs:0.05,0.92) {$(a)$};
\end{axis}
\end{tikzpicture}
\vspace{0.1in}
\begin{tikzpicture}
\begin{axis}[xlabel=$\phi$,ylabel=$\|u\|_\infty$,y label style={rotate=0},width=\textwidth,height=0.4\textwidth,xmin=0,xmax=0.26,ymin=0,ymax=3.05,enlargelimits=0, xticklabel style={/pgf/number format/fixed}, xtick = {0,0.05,0.1,0.15,0.2,0.25}]
\pgfplotstableread{data/periodic2d.txt}\periodictwod;
\addplot[mark=o] table[x=phi,y=umax] {\periodictwod};
\pgfplotstableread{data/randomsummary2d.txt}\randomtwod;
\addplot[mark = square, error bars/.cd, y dir = both, y explicit] table[x = phi, y = umaxmean, y error plus = eplus, y error minus = eminus]{\randomtwod};
\addplot[dashed, no marks]{3/2};
\node at (rel axis cs:0.05,0.92) {$(b)$};
\end{axis}
\end{tikzpicture}
\caption{\label{fig:randper} Peak density $\|u\|_\infty$ of minimizers of the Cahn-Hilliard equation (\ref{eq:CH3}) for two classes of obstacle potentials $W(\vec{x})$. We vary $\phi$, the volume fraction of the computational domain in which the environmental potential $W$ is effectively infinite (here, $10^4$). Elsewhere, $W=0$. These choices of W model, respectively, regions barren and lush with vegetation. Open circles correspond to choosing for $W$ a periodic pattern of square wells, representing engineered cropland with barriers between crop beds. Squares correspond to choosing for $W$ a uniformly random distributed pattern, producing an irregular, clumpy vegetation landscape. Error bars represent maximum and minimum values over ten realizations of $W$ for the random case. Dotted lines represent $\|u\|_\infty = 3/2$, the energetically preferred peak density for large mass solutions in a large domain. (a) Minimizers in one dimension, with domain length $L=100$, population mass $M=50$, simulated with $N=480$ gridpoints. (b) Minimizers in two dimensions, with domain length $L = 36$, population mass $M=648$, and $N=120$ gridpoints along each axis. For both (a) and (b), the average population density is $1/2$. In both panels, a randomly generated $W$ tends to increase peak density $\|u\|_\infty$ above $3/2$ while periodic square wells tend to decrease it below $3/2$.}
\end{figure}

In short, periodic potentials can suppress the peak density below $3/2$, while random potentials can concentrate density at values greater than $3/2$. Figure~\ref{fig:randper} provides a systematic investigation for varying volume fraction $\phi$ in one and two dimensions. In both cases, at low $\phi$, $\|u\|_\infty \approx 3/2$ (dotted line) for both random (shown as squares) and periodic (shown as circles) potentials. As $\phi$ increases, the periodic potential suppresses peak density below $3/2$ whereas the random potential increases it above. This result suggests the viability of carefully engineering resource layout as a strategy for reducing a population's peak density.

%%%%%%%%%%%%%%%%%%%%%%%%%%%%%%%%%%%%%%
%%%%%%%%%%%%%%%%%%%%%%%%%%%%%%%%%%%%%%

%%%%%%%%%%%%%%%%%%%%%%%%%%%%%%%%%%%%%%
%%%%%%%%%%%%%%%%%%%%%%%%%%%%%%%%%%%%%%

\section{Conclusion}

Understanding how populations respond to social forces and environmental cues such as resource distribution is essential for modeling biological groups. For example, for desert locusts, spatially varying resource distributions can concentrate populations, which in the wild can lead to collective gregarization and dangerous outbreaks \cite{DesColSim2000}. We have laid out a framework for studying these problems by exploiting energy minimization of the governing models.

More specifically, we began with aggregation model describing nonlocal social attraction, local, nonlinear repulsion, and an external potential modeling the environment. This model possesses an energy that is minimized by the dynamics. However, the model poses (at least) three challenges. First, time-dependent simulations are costly, and may converge slowly to attractors. Second, attractors typically have compact support. Standard simulation methods may produce oscillations emanating from contact points which typically violates nonnegativity of the solution. Third, the nonlocality couples all spatial locations, rendering numerical computations expensive, especially in two or more dimensions.

To address the first two challenges, we exploit the energy minimizing dynamics by focusing on the analysis and numerical computation of minimizers. This avoids costly computations of dynamics, and additionally, minimization algorithms allow us to enforce nonnegativity of the solution as a simple constraint in the optimization procedure. To address the third challenge, we perform a long-wave approximation of our model to obtain a fourth-order degenerate Cahn-Hilliard equation which is local in space while preserving the energy minimizing character of the original model. In summary, we have developed an approach to studying aggregations that replaces simulating a nonlocal equation with minimizing a local energy subject to a nonnegativity constraint.

In the absence of an external (that is, environmental) potential, minimizers of the nonlocal and Cahn-Hilliard equations agree well at large masses and away from the contact points. In the limit of large domains and modest densities, these minimizers approach a compactly supported clump of population with peak density $3/2$ and steep edges.

An external potential drives mass to accumulate near potential minima, often creating collections of compactly supported clumps. Many environmental landscapes can be partitioned into regions that are either inviting or inhospitable to biological organisms. This characterization arises naturally in numerical computations for sufficiently steep potential wells. Thus, we model these landscapes via obstacle potentials that are effectively infinite (inhospitable) on some subset of the domain and zero (inviting) elsewhere. 

In this obstacle potential testbed, we ask how resource distribution effects peak population density. In the absence of an external potential and at modest average population densities, aggregations form with peak density $3/2$, as mentioned above. A spatially random obstacle potential modeling the natural environment drives peak densities higher, even when the inhospitable portion of the domain is less than $10\%$ of the total. However, a periodic pattern of square wells modeling divided crop beds can reduce the peak density below $3/2$.

Anthropologists have chronicled the use of periodic planting patterns since ancient times, and typically attribute this strategy to better water management and thus increased crop yield. We are intrigued, but admit that it is pure speculation, to ask whether this strategy might confer any pest control advantages. Regardless, the biological literature notes that for locusts, ``fractal dimension'' of resources in the environment can drive large peak population densities \cite{DesColSim2000}. While perhaps just a caricature, our simple model suggests that carefully distributing resources can inhibit populations from reaching their intrinsic peak density and plausibly could help control undesirable biological phenomena such as gregarious locust outbreaks that are triggered by surpassing a density threshold.

\section*{Acknowledgments}
AJB thanks Macalester College for hosting him during much of this work, and is grateful to Harvey Mudd College for their generous travel support. We are also grateful to Alese Halvorson and Iris Vrioni, who contributed to exploratory stages of this work during their time as undergraduate students at Macalester College. We thank the two anonymous referees for their insightful and constructive comments.

\bibliography{master_bibliography}
\bibliographystyle{siam}

\end{document}